\documentclass[letterpaper]{article} 
\usepackage{aaai2026}  
\usepackage{times}  
\usepackage{helvet}  
\usepackage{courier}  
\usepackage[hyphens]{url}  
\usepackage{graphicx} 
\usepackage{natbib}  
\usepackage{caption} 
\usepackage{algorithm}
\usepackage{algorithmic}
\usepackage{booktabs} 
\usepackage{adjustbox}
\usepackage[table]{xcolor}
\usepackage{amsfonts}
\usepackage{amsmath}
\usepackage{multirow}
\usepackage{adjustbox}

\usepackage{newfloat}
\usepackage{listings}
\DeclareCaptionStyle{ruled}{labelfont=normalfont,labelsep=colon,strut=off} 
\floatstyle{ruled}
\newfloat{listing}{tb}{lst}{}
\floatname{listing}{Listing}
\title{Dual-Path Knowledge-Augmented Contrastive Alignment Network for Spatially Resolved Transcriptomics}
\author{
    Wei Zhang\equalcontrib, 
    Jiajun Chu\equalcontrib, 
    Xinci Liu,
    Chen Tong, 
    Xinyue Li\thanks{Corresponding author.}
}
\affiliations{
    Department of Data Science, College of Computing, City University of Hong Kong, Hong Kong SAR, China\\
    xinyueli@cityu.edu.hk
}

\usepackage{bibentry}

\begin{document}

\maketitle

\begin{abstract}
Spatial Transcriptomics (ST) is a technology that measures gene expression profiles within tissue sections while retaining spatial context. It reveals localized gene expression patterns and tissue heterogeneity, both of which are essential for understanding disease etiology. However, its high cost has driven efforts to predict spatial gene expression from whole slide images. Despite recent advancements, current methods still face significant limitations, such as under-exploitation of high-level biological context, over-reliance on exemplar retrievals, and inadequate alignment of heterogeneous modalities. To address these challenges, we propose \textbf{DKAN}, a novel \textbf{D}ual-path \textbf{K}nowledge-\textbf{A}ugmented contrastive alignment \textbf{N}etwork that predicts spatially resolved gene expression by integrating histopathological images and gene expression profiles through a biologically informed approach. Specifically, we introduce an effective gene semantic representation module that leverages the external gene database to provide additional biological insights, thereby enhancing gene expression prediction. Further, we adopt a unified, one-stage contrastive learning paradigm, seamlessly combining contrastive learning and supervised learning to eliminate reliance on exemplars, complemented with an adaptive weighting mechanism. Additionally, we propose a dual-path contrastive alignment module that employs gene semantic features as dynamic cross-modal coordinators to enable effective heterogeneous feature integration. Through extensive experiments across three public ST datasets, DKAN demonstrates superior performance over state-of-the-art models, establishing a new benchmark for spatial gene expression prediction and offering a powerful tool for advancing biological and clinical research.
\end{abstract}

\begin{links}
    \link{Code}{https://github.com/coffeeNtv/DKAN}
\end{links}

\section{Introduction}

\begin{figure}[htbp]
  \centering
  \includegraphics[width=\linewidth]{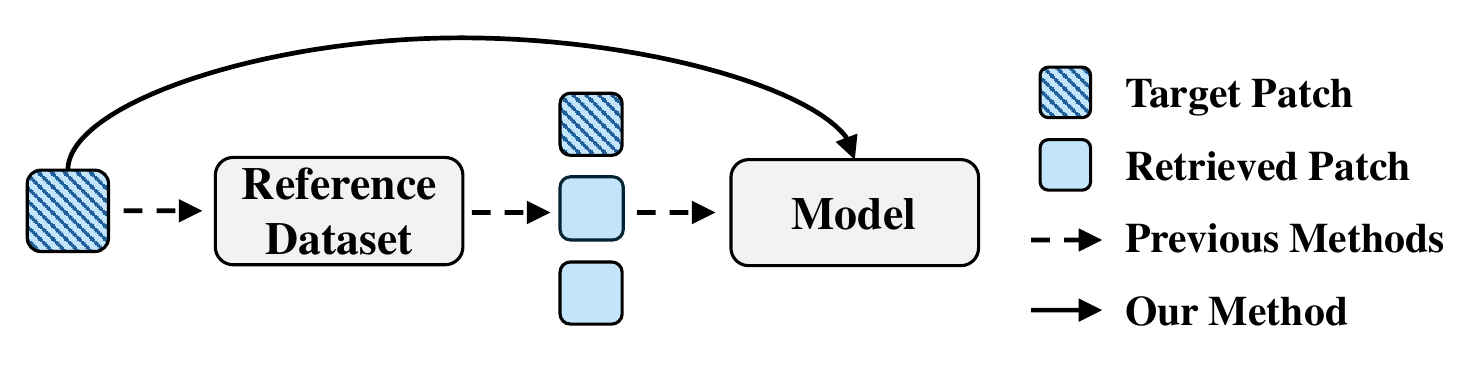}
  \caption{Pipeline comparison. Existing contrastive learning and exemplar-guided methods  require constructing a reference dataset and retrieving similar patches as intermediate steps, whereas our one-stage contrastive learning method operates through a straightforward pipeline.}
  \label{fig:teaser}
\end{figure}

Spatial Transcriptomics (ST) is an advanced technology that measures gene expression profiles within tissue sections while preserving their spatial context, often by integrating data with Whole Slide Images (WSIs)~\cite{sci}. This spatially resolved perspective is pivotal in revealing the heterogeneity of gene expression across tissue microenvironments, offering critical insights into developmental processes, disease progression, and cell-cell interactions. 

Despite the transformative potential of ST techniques, they still face limitations, including relatively low resolution, typically at the multicellular level, and high technical costs, which hinder their broader adoption~\cite{moses_museum_2022}. In contrast, Hematoxylin and Eosin (H\&E) stained WSIs, as the gold standard in pathology, offer a cost-effective and widely accessible alternative. Their widespread availability and low costs make them suitable for supporting numerous downstream tasks such as survival prediction~\cite{zhang2024samamba,zhang2025integrating}, stain transfer~\cite{li2023adaptive,zhang2024ppt} and especially spatial transcriptomics~\cite{stalign,zhang2025bridging}. The potential of WSIs to predict spatially resolved gene expression has been successfully demonstrated~\cite{stnet, he2rna}, utilizing morphological and spatial details. More recently, several models have expanded on this foundation, further improving prediction accuracy through innovative approaches to leverage the rich tissue information in WSIs~\cite{triplex, m2ost}.

Current approaches for spatial gene expression prediction predominantly exploit the rich spatial information embedded in WSIs, extracting image features at various levels, including local~\cite{bleep,sepal}, global~\cite{histogene,hist2st}, and multi-scale representations~\cite{m2ost,stalign}. Furthermore, several models incorporate multimodal contrastive learning to align imaging data with gene expression profiles within a shared low-dimensional embedding space ~\cite{bleep,mclstexp}. This alignment enables the models to effectively capture the intricate relationships between image-derived features and gene expression patterns.

Despite recent advancements, several challenges persist. First, many models rely heavily on image features derived from pixel intensity (e.g., color distribution) and cellular structure (e.g., shape and texture)~\cite{stnet,histogene}. While these low-level visual cues are informative, they often fail to capture high-level semantic information, such as gene functions, biological pathways, or disease associations, limiting the depth of biological interpretation. Second, one notable challenge lies in the inclusion of additional and potentially redundant steps in models based on contrastive learning and exemplar-guided strategies. As shown in Figure~\ref{fig:teaser}, these pipelines typically involve constructing a reference dataset from all patches in the training set, retrieving similar patches, and feeding both the retrieved patches and the target patch into the model. While effective, this multi-step process introduces complexity that may not always be necessary~\cite{bleep,mclstexp,stalign}. Streamlining such workflows into a more cohesive approach, particularly in constrained settings or with limited datasets, remains challenging. Lastly, while existing methods leverage multi-scale image features~\cite{hist2st,triplex} or incorporate auxiliary modalities~\cite{sgn} to address modality-specific semantics, their fusion strategies often fail to adequately preserve biologically relevant interactions. This limitation constrains performance, a gap further exacerbated by the absence of frameworks that explicitly incorporate gene functional semantics into multimodal alignment.

To address these challenges, we propose DKAN, a novel \textbf{D}ual-path \textbf{K}nowledge-\textbf{A}ugmented contrastive alignment \textbf{N}etwork for spatial gene expression prediction. Unlike previous methods, DKAN integrates gene functional semantics into contrastive learning, enabling the biologically grounded fusion of histopathological images and expression profiles through a unified one-stage paradigm. Our major contributions are summarized as follows:
\begin{itemize}

\item We propose a novel paradigm for spatial gene expression prediction by incorporating gene functional semantics into contrastive learning, enabling the model to capture high-level biological context beyond low-level image features and align predictions with established genomic knowledge.
    
\item We develop a unified one-stage contrastive learning framework that integrates supervised and contrastive objectives via adaptive weighting, simplifying the pipeline by removing exemplar dependence and eliminating separate storage or retrieval steps.

\item We introduce a dual-path contrastive alignment module that processes image and gene expression features separately, avoiding forced direct alignment of heterogeneous modalities. Leveraging gene semantics enables precise multimodal integration into a shared embedding space, overcoming limitations of prior fusion strategies.

\item We conduct extensive experiments on three public ST datasets, demonstrating DKAN consistently outperforms State-Of-The-Art (SOTA) models across benchmarks.
\end{itemize}

\begin{figure*}[htbp]
  \centering
  \includegraphics[width=\linewidth]{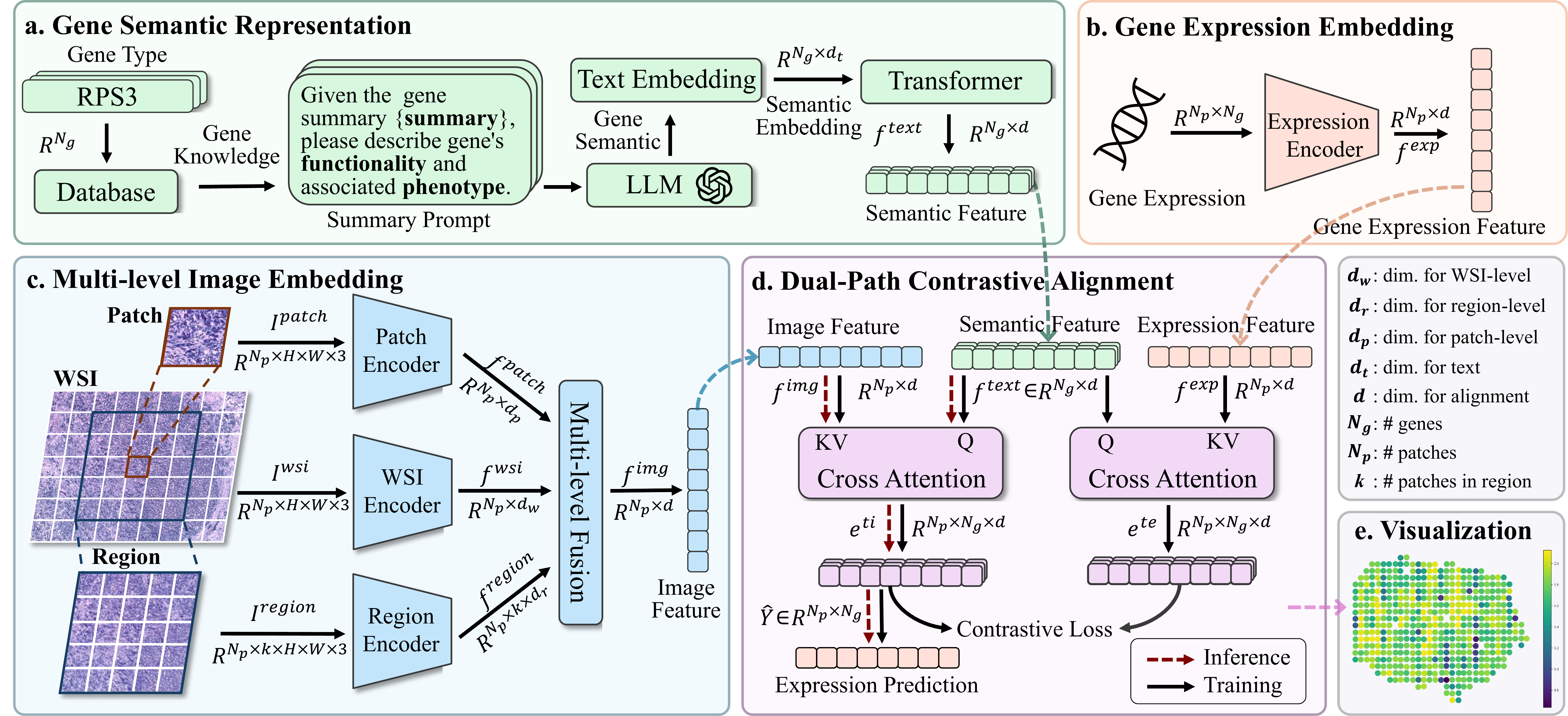}
  \caption{Overview of DKAN framework. (a) Gene semantic feature representation module. (b) Gene expression feature embedding module. (c) Multi-level image embedding module. (d) Dual-path contrastive alignment module. (e) Example spatial gene expression map for FN1 in the STNET dataset.}
  \label{fig:framework}
\end{figure*}

\section{Related Work}
\subsection{Spatial Gene Expression Prediction}
\label{sec:work_prediction}
Spatial gene expression prediction seeks to model gene activity from WSIs by capturing both visual and spatial features. Existing methods fall into three main categories:

\textbf{Local Methods.} Local approaches focus on the target patch and its immediate surroundings~\cite{egn,eggn,sgn,bleep,mclstexp}. ST-Net~\cite{stnet} uses a pretrained DenseNet-121~\cite{densenet} to extract patch-level features for prediction. EGN~\cite{egn} enhances patch representations through image reconstruction and exemplar retrieval. EGGN~\cite{eggn} extends EGN by applying graph convolutional networks to model relationships between the patch and its exemplars. SEPAL~\cite{sepal} constructs a neighborhood graph and applies a graph neural network to capture local dependencies. These methods prioritize localized information and may not account for the broader spatial context.

\textbf{Global Methods.} Global methods incorporate positional and contextual information across the entire WSI~\cite{thitogene,sgn}. HisToGene~\cite{histogene} uses vision transformers~\cite{vit} with positional encoding to model inter-patch relationships. HE2RNA~\cite{he2rna} clusters patches into supertiles and aggregates them to form global contextual features. THItoGene~\cite{thitogene} extracts deep molecular features using dynamic convolution and capsule modules, integrates them with positional data via ViT, and refines predictions using a graph attention network. Unlike local methods, global models leverage the full image context for more informed predictions.

\textbf{Multi-Scale Methods.} Multi-scale methods capture biological patterns at various resolutions~\cite{hist2st,triplex,m2ost,stalign}. TRIPLEX~\cite{triplex} combines features from multiple views. M2OST~\cite{m2ost} decouples intra- and inter-scale feature extraction for many-to-one spatial prediction. ST-Align~\cite{stalign} clusters patches into niche-level groups to integrate both local and regional contexts. These methods aim to balance the fine granularity of local models with the broader context provided by global ones.

\subsection{Contrastive Representation Learning}
\label{sec:work_contrastive}

Contrastive learning is a self-supervised method that learns discriminative representations by pulling similar pairs together and pushing dissimilar pairs apart~\cite{cpc}. In spatial gene expression tasks, contrastive learning aligns visual and transcriptomic modalities in a shared embedding space~\cite{bleep,mclstexp,stalign}. BLEEP~\cite{bleep}, inspired by CLIP~\cite{clip}, embeds images and gene profiles jointly to enable retrieval-based inference. mclSTExp~\cite{mclstexp} refines this by encoding gene expression with learnable position embeddings for better spatial integration. ST-Align~\cite{stalign} advances contrastive learning by curating 1.3 million image-gene pairs and introducing a multi-scale feature extractor with a three-target alignment strategy. This enhances the model’s ability to capture complex structural patterns in spatial transcriptomics.

\section{Methodology}

\subsection{Problem Formulation}
\label{sec:method_problem}
Spatial gene expression prediction is framed as a regression task. Given $N_p$ image patches extracted from a WSI, represented as $X \in \mathbb{R}^{N_p \times H \times W \times 3}$, where $H$ and $W$ are the height and width of each patch, the goal is to predict gene expression levels. A learnable mapping function $f$ is applied to produce predictions $\hat{Y} = f(X) \in \mathbb{R}^{N_p \times N_g}$, where $N_g$ denotes the number of target genes.

\subsection{Overview}
\label{sec:method_overview}
As illustrated in Figure~\ref{fig:framework}, our framework integrates high-level gene semantics by retrieving information from an external gene database~\cite{ncbi_database_2024} and leveraging prompts to tap into the summarization capabilities and domain knowledge of large language models (LLMs). To extract informative visual features from WSIs, we adopt a multi-scale strategy that captures representations at the patch, region, and whole-slide levels. These features are then fused to form a comprehensive visual embedding. To effectively align gene expression, image, and textual modalities, we introduce a dual-path knowledge-augmented contrastive alignment module, which employs two distinct contrastive pathways for robust multimodal integration.

\subsection{Gene Semantic Representation}
\label{sec:method_text}

As shown in Figure \ref{fig:framework}(a), for $N_g$ genes of interest, we designed a workflow to extract semantic features $f^{text}$. We retrieved gene-related knowledge from a well-established gene database NCBI~\cite{ncbi_database_2024}. However, the retrieved gene knowledge lacks structural uniformity, with some information being redundant or incomplete. To address this issue, we leveraged the summarization capabilities and embedded knowledge of LLM (GPT-4o) to generate accurate and efficient gene semantic texts. 

Specifically, we embedded the gene knowledge into a prompt, which includes role definitions, task requirements, and output specifications, before feeding it into the LLM to produce the gene semantic text. The prompt is provided in the supplementary materials. Subsequently, we employed BioBERT~\cite{biobert} as our text embedding model to extract textual features, generating semantic embeddings of dimensionality $d_t=1024$. This model, pre-trained on the extensive biomedical semantic corpus, excels at capturing domain-specific contextual representations effectively. The semantic features are processed by a standard transformer module, preceded by a linear projection to ensure dimensional alignment for multimodal fusion. The transformer module efficiently captures global dependencies and commonalities among semantic embeddings, ultimately yielding the final semantic features $f^{text}$.

\subsection{Gene Expression Embedding}
\label{sec:method_expression}
As illustrated in Figure \ref{fig:framework}(b), the gene expression with shape ${N_p}\times{N_g}$ is processed by the gene expression encoder to generate gene expression features $f^{exp}\in \mathbb{R}^{{N_p}\times{d}}$, ensuring feature dimension consistency between gene semantic and image features. Specifically, the expression encoder first projects the input to a $d$-dimensional space via a linear layer, applies GELU activation, and then processes features through a second linear layer with dropout. To stabilize gradient flow, we employ residual connections: the initial linear projection's output is directly added to the final dropout output, followed by layer normalization for feature standardization. This design mitigates gradient vanishing while maintaining feature discriminability.

\subsection{Multi-level Image Embedding}
\label{sec:method_image}
Given the large size of WSIs, relying solely on either WSI-level images ($I^{wsi}$) or patch-level images ($I^{patch}$) is insufficient to fully capture their morphological complexity. While WSIs offer rich global context, a significant gap remains between the global view and the localized detail at the patch level. To bridge this gap, as illustrated in Figure~\ref{fig:framework}(c), we introduce a region-level representation ($I^{region}$) by selecting the k nearest neighboring patches around each target patch.

Our model extracts image patches from the WSI at these three hierarchical levels and processes them using dedicated encoders. Specifically, for the WSI-level features $f^{wsi}$ and the region-level features $f^{region}$, we utilize UNI~\cite{uni}, a general-purpose foundation model pre-trained on extensive WSI datasets for computational pathology. Due to the scale constraints and computational demands of WSIs and region-level images, UNI serves as a fixed feature extractor without updating its weights during training. To enhance feature adaptability, we append a multi-head transformer after each UNI encoder. For the patch-level feature $f^{patch}$, we employ ResNet18~\cite{resnet18} as the encoder. To adapt it to feature extraction, we remove the final pooling and fully connected layers, retaining only the activations of the last hidden layer as the output. Notably, the parameters of ResNet18 remain trainable. 

To effectively integrate multi-scale features, we employ two cross-attention mechanisms: one fuses the WSI and the region-level images, while the other combines the WSI and the patch-level images, with WSI-level features serving as the query in both cases. The resulting fused features from these two groups are then summed to produce the final multi-scale feature of the image $f^{img}$.

\begin{table*}[htb]
\small
\centering
\setlength{\tabcolsep}{5pt}
\begin{adjustbox}{width=0.99\textwidth}
\begin{tabular}{c|c|l|cccccc}
\toprule 

\multicolumn{3}{c|}{Comparison Settings} & \multicolumn{2}{c}{Error} & \multicolumn{4}{c}{PCC} \\
\cmidrule(lr){0-2} 
\cmidrule(lr){4-5}
\cmidrule(lr){6-9}
& Type & Model & \multicolumn{1}{c}{MAE$\downarrow$} & \multicolumn{1}{c}{MSE$\downarrow$} & \multicolumn{1}{c}{ALL$\uparrow$} & \multicolumn{1}{c}{HPG$\uparrow$} & \multicolumn{1}{c}{HEG$\uparrow$} & \multicolumn{1}{c}{HVG$\uparrow$}\\ 
\midrule
\multirow{13}{*}{\rotatebox{90}{HER2+ Dataset}} 
& \multirow{4}{*}{Local} 
& ST-Net &$0.432\pm0.05$&$0.311\pm0.07$&$0.150\pm0.13$&$0.287\pm0.19$&$0.115\pm0.11$&$0.090\pm0.08$\\

& & BLEEP &$0.401\pm0.03$&$0.277\pm0.05$&$0.151\pm0.11$&$0.277\pm0.16$
&$0.246\pm0.09$&$0.261\pm0.07$\\

& & EGN &$0.366\pm0.04$&$0.229\pm0.05$&$0.204\pm0.12$&$0.364\pm0.16$&$0.152\pm0.09$&$0.120\pm0.05$\\

& & mclSTExp &$0.398\pm0.04$&$0.272\pm0.05$&$0.163\pm0.11$&$0.289\pm0.16$&$0.114\pm0.08$&$0.091\pm0.06$\\

\cmidrule(lr){2-9}
&\multirow{3}{*}{Global}
& HisToGene &$0.388\pm0.06$&$0.253\pm0.07$&$0.150\pm0.09$&$0.295\pm0.15$&$0.099\pm0.07$&$0.079\pm0.05$\\

& & THItoGene &$0.424\pm0.05$&$0.291\pm0.06$&$0.051\pm0.05$&$0.118\pm0.08$&$0.045\pm0.05$&$0.030\pm0.03$\\

& & SGN &$0.734\pm0.20$&$0.749\pm0.38$&$0.035\pm0.03$&$0.065\pm0.05$&$0.022\pm0.03$&$0.017\pm0.03$\\

\cmidrule(lr){2-9}

& \multirow{4}{*}{Multi-view}
& Hist2ST &$0.417\pm0.07$&$0.293\pm0.08$&$0.193\pm0.10$&$0.360\pm0.17$&$0.126\pm0.07$&$0.109\pm0.03$\\

& & TRIPLEX &$0.364\pm0.05$&$0.234\pm0.06$&$0.304\pm0.14$&$0.491\pm0.18$&$0.271\pm0.10$&$0.260\pm0.06$\\

& & M2OST &$0.446\pm0.10$&$0.340\pm0.15$&$0.147\pm0.12$&$0.313\pm0.19$&$0.098\pm0.09$&$0.090\pm0.06$\\

& & \cellcolor{gray!20}DKAN(Ours) &\cellcolor{gray!20}$\textbf{0.361}\pm\textbf{0.04}$&\cellcolor{gray!20}$\textbf{0.224}\pm\textbf{0.06}$&\cellcolor{gray!20}$\textbf{0.330}\pm\textbf{0.13}$&\cellcolor{gray!20}$\textbf{0.531}\pm\textbf{0.15}$&\cellcolor{gray!20}$\textbf{0.317}\pm\textbf{0.09}$&\cellcolor{gray!20}$\textbf{0.304}\pm\textbf{0.07}$
\\
\midrule
\multirow{13}{*}{\rotatebox{90}{STNET Dataset}} 
& \multirow{4}{*}{Local} 
& ST-Net &$0.357\pm0.04$&$0.222\pm0.05$&$0.081\pm0.05$&$0.192\pm0.09$&$0.026\pm0.05$&$0.096\pm0.04$\\

& & BLEEP &$0.369\pm0.02$&$0.235\pm0.02$&$0.095\pm0.05$&$0.193\pm0.10$
&$0.063\pm0.03$&$0.111\pm0.05$\\

& & EGN &$0.354\pm0.02$&$0.214\pm0.03$&$0.107\pm0.05$&$0.207\pm0.09$&$0.089\pm0.04$&$0.108\pm0.04$\\

& & mclSTExp &$0.350\pm0.02$&$0.210\pm0.02$&$0.095\pm0.05$&$0.202\pm0.09$&$0.052\pm0.04$&$0.088\pm0.03$\\

\cmidrule(lr){2-9}
&\multirow{3}{*}{Global}
& HisToGene &$0.326\pm0.02$&$0.180\pm0.02$&$0.103\pm0.04$&$0.217\pm0.11$&$0.060\pm0.02$&$0.074\pm0.03$\\

& & THItoGene &$0.347\pm0.03$&$0.200\pm0.04$&$0.040\pm0.02$&$0.092\pm0.02$&$0.025\pm0.02$&$0.028\pm0.02$\\

& & SGN &$1.180\pm1.78$&$4.952\pm12.00$&$0.027\pm0.01$&$0.048\pm0.02$&$0.014\pm0.01$&$0.027\pm0.01$\\

\cmidrule(lr){2-9}

& \multirow{4}{*}{Multi-view}
& Hist2ST &$0.352\pm0.02$&$0.208\pm0.03$&$0.142\pm0.04$&$0.268\pm0.09$&$0.094\pm0.03$&$0.122\pm0.03$\\

& & TRIPLEX &$0.342\pm0.02$&$0.200\pm0.02$&$0.194\pm0.07$&$0.344\pm0.10$&$0.160\pm0.06$&$0.224\pm0.07$\\

& & M2OST &$0.369\pm0.04$&$0.226\pm0.04$&$0.022\pm0.02$&$0.081\pm0.04$&$0.008\pm0.03$&$-0.001\pm0.03$\\

& & \cellcolor{gray!20}DKAN(Ours) &\cellcolor{gray!20}$\textbf{0.322}\pm\textbf{0.02}$&\cellcolor{gray!20}$\textbf{0.179}\pm\textbf{0.02}$&\cellcolor{gray!20}$\textbf{0.219}\pm\textbf{0.07}$&\cellcolor{gray!20}$\textbf{0.387}\pm\textbf{0.09}$&\cellcolor{gray!20}$\textbf{0.200}\pm\textbf{0.06}$&\cellcolor{gray!20}$\textbf{0.244}\pm\textbf{0.07}$\\

\midrule

\multirow{13}{*}{\rotatebox{90}{cSCC Dataset}} 
& \multirow{4}{*}{Local} 
& ST-Net &$0.410\pm0.05$&$0.262\pm0.07$&$0.170\pm0.08$&$0.289\pm0.09$&$0.079\pm0.05$&$0.065\pm0.02$\\

& & BLEEP &$0.430\pm0.04$&$0.297\pm0.05$&$0.269\pm0.07$&$0.396\pm0.08$
&$0.266\pm0.09$&$0.250\pm0.10$\\

& & EGN &$0.438\pm0.05$&$0.303\pm0.06$&$0.278\pm0.06$&$0.388\pm0.06$&$0.194\pm0.06$&$0.180\pm0.06$\\

& & mclSTExp &$0.445\pm0.05$&$0.311\pm0.06$&$0.168\pm0.04$&$0.291\pm0.08$&$0.098\pm0.03$&$0.096\pm0.05$\\

\cmidrule(lr){2-9}
&\multirow{3}{*}{Global}
& HisToGene &$0.441\pm0.06$&$0.297\pm0.07$&$0.178\pm0.10$&$0.319\pm0.14$&$0.099\pm0.05$&$0.094\pm0.04$\\

& & THItoGene &$0.495\pm0.11$&$0.380\pm0.16$&$0.040\pm0.05$&$0.101\pm0.06$&$0.014\pm0.03$&$0.020\pm0.03$\\

& & SGN &$0.832\pm0.13$&$0.897\pm0.27$&$0.059\pm0.02$&$0.086\pm0.03$&$0.048\pm0.02$&$0.044\pm0.01$\\

\cmidrule(lr){2-9}

& \multirow{4}{*}{Multi-view}
& Hist2ST &$0.468\pm0.11$&$0.338\pm0.16$&$0.185\pm0.14$&$0.261\pm0.15$&$0.133\pm0.09$&$0.110\pm0.07$\\

& & TRIPLEX &$0.415\pm0.06$&$0.278\pm0.08$&$0.363\pm0.07$&$0.476\pm0.07$&$0.276\pm0.07$&$0.272\pm0.06$\\

& & M2OST &$0.443\pm0.07$&$0.313\pm0.09$&$-0.018\pm0.04$&$0.126\pm0.09$&$0.012\pm0.07$&$-0.041\pm0.04$\\

& & \cellcolor{gray!20}DKAN(Ours) &\cellcolor{gray!20}$\textbf{0.383}\pm\textbf{0.05}$&\cellcolor{gray!20}$\textbf{0.239}\pm\textbf{0.06}$&\cellcolor{gray!20}$\textbf{0.407}\pm\textbf{0.08}$&\cellcolor{gray!20}$\textbf{0.508}\pm\textbf{0.08}$&\cellcolor{gray!20}$\textbf{0.346}\pm\textbf{0.09}$&\cellcolor{gray!20}$\textbf{0.321}\pm\textbf{0.08}$  \\
\bottomrule
\end{tabular}
\end{adjustbox}
\caption{Comparison with SOTA methods. The best results are highlighted in bold.}
\label{tab:baseline}
\end{table*}

\subsection{Dual-Path Contrastive Alignment}
\label{sec:method_fusion}
After extracting the image, semantic, and expression features, we propose a novel dual-path knowledge-augmented contrastive alignment paradigm for multimodal alignment as shown in Figure \ref{fig:framework}(d). Our approach leverages gene semantic features as dynamic cross-modal coordinators, operating through two parallel pathways. In the image pathway, gene semantic knowledge serves as a ``functional query instruction'' to filter morphology-related regions from image features. Similarly, in the expression pathway, gene semantic knowledge acts as a ``distribution correction factor'' to constrain the predicted gene expression features, ensuring alignment with the established biological pathway logic. In the implementation, we employ a cross-attention module, using the semantic feature as the query. Each semantic feature independently queries the image and expression features, ultimately generating gene knowledge-augmented representations, denoted as $e^{ti}$ and $e^{te}$ respectively.

Inspired by CLIP~\cite{clip}, we apply contrastive learning to align $e^{ti}$ and $e^{te}$ in the latent embedding space. One distinctive aspect of this method is that, instead of forcing the alignment of the heterogeneous image and gene expression modalities directly, each modality interacts independently with the gene semantic knowledge, achieving implicit alignment through knowledge-guided queries. The decoupling of the image and gene expression modules enhances flexibility and reduces inter-modal dependencies.

To eliminate the dependency on exemplars and streamline the workflow, we adopt a unified one-stage framework for contrastive learning and seamlessly integrate it with supervised training. In the training phase, all modalities are utilized, whereas during inference, only the image and semantic modalities are used. Consequently, the loss function combines a contrastive loss and a supervised loss. The contrastive loss is indicated in Equation~\ref{eq:contrast}, where positive samples are representations of the same gene paired together, and negative samples are drawn from representations of different genes. Here $sim(\cdot,\cdot)$ denotes the cosine similarity function that measures the alignment between feature vectors, and $\tau$ is a temperature parameter that controls the sharpness of the similarity distribution.

\begin{equation}
\mathcal{L}_{cont} = -\sum_{i} \log \frac{\exp(sim(e_{ti}^i, e_{te}^i) / \tau)}{\sum_{j} \exp(sim(e_{ti}^i, e_{te}^j) / \tau)}.
\label{eq:contrast}
\end{equation}

For the supervised loss, we calculate the mean squared error (MSE) between the predicted gene expression \(\hat{Y}\) and the ground truth gene expression \(Y\). This can be further enhanced by knowledge distillation~\cite{triplex}, which improves prediction consistency and generalization by aligning intermediate representations with the final output. These intermediate predictions, \(\hat{Y}_{img}\), \(\hat{Y}_{patch}\), \(\hat{Y}_{wsi}\), and \(\hat{Y}_{region}\), are obtained through linear transformations of the model features \(f^{img}\), \(f^{patch}\), \(f^{wsi}\), and \(f^{region}\). To enforce both accuracy and coherence, we compute the MSE between these intermediate predictions and two targets: the ground truth \(Y\) and the final predicted output \(\hat{Y}\), with their contributions balanced by a hyperparameter \(\lambda\).

The distillation-aware supervised loss for each intermediate prediction $d \in \mathcal{D}$ is defined in Equation~\ref{eq:kd}:

\begin{equation}
\mathcal{L}_{d} = \lambda \|\hat{Y}_d - \hat{Y}\|^2 + (1 - \lambda) \|\hat{Y}_d - Y\|^2,
\label{eq:kd}
\end{equation}

where $\mathcal{D} = \{\text{img}, \text{patch}, \text{wsi}, \text{region}\}$. The total supervised loss is then aggregated across all intermediate predictions, combined with the MSE between the ground truth \(Y\) and the final predicted output \(\hat{Y}\):

\begin{equation}
\mathcal{L}_{sup} = \sum_{d\in{\mathcal{D}}}{\mathcal{L}_d}+\|\hat{Y} - Y\|^2.
\end{equation}

To ensure balanced optimization between the supervised loss ($\mathcal{L}_{sup}$) and the contrastive loss ($\mathcal{L}_{cont}$) which exhibit different numerical scales and convergence characteristics, we propose an adaptive weighting scheme. The weights are dynamically adjusted based on the real-time loss values to maintain appropriate gradient contributions from each objective. Specifically, the weighting coefficients are computed as the normalized reciprocals of the respective losses, to ensure that the loss with a smaller value could receive a higher weight, allowing the model to dynamically prioritize the more reliable objective during training. The final composite loss function is thus formulated as:

\begin{equation}
\mathcal{L} = w_{sup} \mathcal{L}_{sup} + w_{cont}\mathcal{L}_{cont}.
\end{equation}

\section{Experiments}

\subsection{Datasets}
\label{sec:exp_datasets}
We evaluated our approach on three public ST datasets: two human breast cancer (BC) datasets and one cutaneous squamous cell carcinoma (cSCC) dataset. The HER2+ BC dataset~\cite{andersson_her2_2021} includes 36 samples from 8 patients, with 13,620 spots and 14,873 genes profiled per spot. The STNET BC dataset~\cite{stnet} contains 68 WSIs from 23 patients, comprising 30,612 spots and 26,949 genes per spot. The cSCC dataset~\cite{andrew_skin_2020} consists of 12 samples from 4 patients, with 8,671 spots and 17,047 genes measured per spot.

\begin{table}[htb]
\small
  \setlength{\tabcolsep}{4pt}
  \centering
  \begin{tabular}{l|cccccc}
    \toprule
     & \multicolumn{2}{c}{Error} & \multicolumn{4}{c}{PCC} \\
    \cmidrule(lr){2-3}
    \cmidrule(lr){4-7}
    Text Emb. &MAE$\downarrow$&MSE$\downarrow$&ALL$\uparrow$&HPG$\uparrow$ &HEG$\uparrow$ &HVG$\uparrow$\\
    \midrule
    
    Conch&0.324&0.181&0.211&0.374&0.194&0.230\\
    PLIP&0.323&\textbf{0.179}&0.217&0.381&0.194&\textbf{0.244}\\
    BioGPT&0.327&0.183&0.217&0.385&0.193&0.239\\
    \rowcolor{gray!20}BioBERT&\textbf{0.322}&\textbf{0.179}&\textbf{0.219}&\textbf{0.387}&\textbf{0.200}&\textbf{0.244}\\
    \midrule
    Image Emb. &MAE$\downarrow$&MSE$\downarrow$&ALL$\uparrow$&HPG$\uparrow$ &HEG$\uparrow$ &HVG$\uparrow$\\
    \midrule
    ResNet18&0.341&0.197&0.202&0.347&0.176&0.224\\
    ResNet50&0.354&0.212&0.196&0.349&0.167&0.215\\
    PLIP&0.333&0.188&0.190&0.333&0.157&0.211\\
    Conch&0.324&\textbf{0.177}&0.209&0.369&0.179&0.231\\
    \rowcolor{gray!20}UNI&\textbf{0.322}&0.179&\textbf{0.219}&\textbf{0.387}&\textbf{0.200}&\textbf{0.244}\\
  \bottomrule
\end{tabular}
\caption{Ablation studies on text and image encoders.}
\label{tab:embedding_stnet}
\end{table}

\subsection{Evaluation and Metrics}
\label{sec:exp_evaluation}
To ensure robust evaluation, we applied cross-validation strategies tailored to each dataset, ensuring no patient overlap between training and test sets. For STNET, we used 8-fold cross-validation. For the smaller HER2+ and cSCC datasets, we adopted leave-one-patient-out cross-validation, with 8 folds for HER2+ and 4 folds for cSCC, where each fold used one patient's samples for testing and the rest for training. This setup aligns with prior work~\cite{triplex} to ensure fair comparison.

We evaluated our model using six metrics to ensure comprehensive assessment and comparability with prior studies~\cite{egn,bleep,triplex}. These include Mean Absolute Error (MAE), Mean Squared Error (MSE), and Pearson Correlation Coefficient (PCC) across: (1) all genes of interest, (2) the top 50 Highly Predictive Genes (HPG), (3) the top 50 Highly Expressed Genes (HEG), and (4) the top 50 Highly Variable Genes (HVG). PCC was computed per gene across all spots within each sample and averaged over all cross-validation folds.

\subsection{Implementations} 
To align with previous studies~\cite{stnet,triplex}, all patches were segmented with dimensions of $H$=$W$=224 pixels, and regions were constructed using $k$=25 (a 5×5 patch grid). We select $N_g$=250 spatially variable genes for training to align with previous studies. We use the Adam optimizer with a learning rate of 0.0001 and a StepLR scheduler (step size=50, gamma=0.9). The temperature $\tau$ in the contrastive loss was set to 0.1 for HER2+ and STNET and 0.08 for cSCC. Image encoders included UNI for WSI and region levels ($d_h$=$d_r$=1024) and ResNet18 for patch level ($d_p$=512), while the text embedding model, BioBERT, produced embeddings with $d_t$=1024. All models were trained on a NVIDIA RTX A800 GPU with a batch size of 128.

\begin{figure}[htbp]
  \centering
  \includegraphics[width=\linewidth]{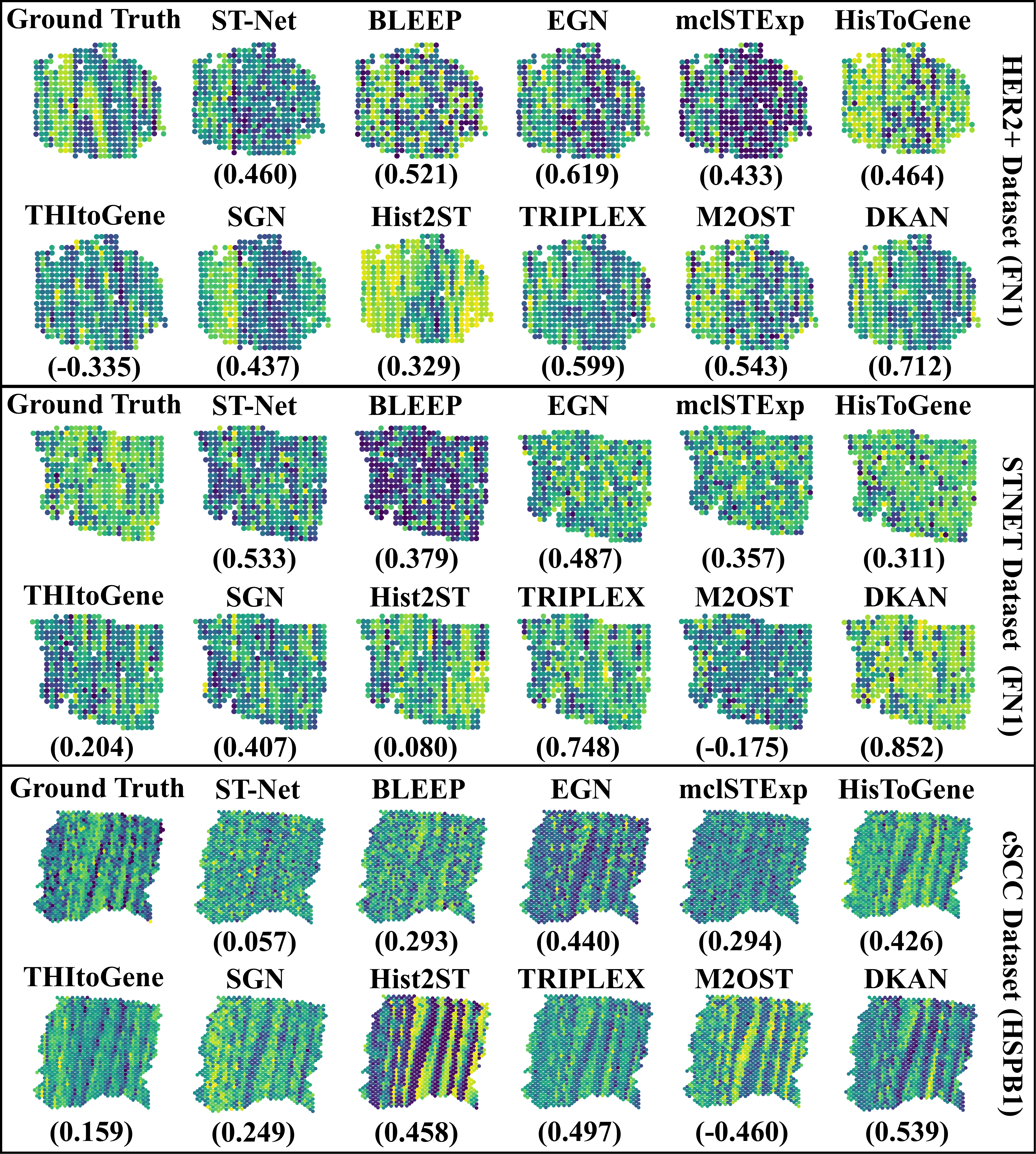}
  \caption{Visualization of expression patterns of cancer biomarker genes alongside PCC values for all datasets.  }
  \label{fig:visual_main}
\end{figure}

\subsection{Experimental Results and Analysis}
\label{sec:exp_results}
We compared our proposed model, DKAN, against extensive SOTA baselines across three categories: (1) Local methods: ST-Net~\cite{stnet}, BLEEP~\cite{bleep}, EGN~\cite{egn}, and mclSTExp~\cite{mclstexp}; (2) Global methods: HisToGene~\cite{histogene}, THItogene~\cite{thitogene}, and SGN~\cite{sgn}; and (3) Multi-scale methods: Hist2ST~\cite{hist2st}, TRIPLEX~\cite{triplex}, and M2OST~\cite{m2ost}. As shown in Table~\ref{tab:baseline}, DKAN consistently outperforms all baselines across datasets and evaluation metrics.

Take the HER2+ dataset as an example, DKAN achieves the lowest MAE (0.361) and MSE (0.224), along with the highest PCC values for all genes (0.330), HPG (0.531), HEG (0.317), and HVG (0.304). In comparison, the current SOTA method TRIPLEX reports 0.364 (MAE), 0.234 (MSE), and PCCs of 0.304 (all genes), 0.491 (HPG), 0.271 (HEG), and 0.260 (HVG). These results demonstrate DKAN’s superior performance over TRIPLEX and other baselines. Notably, DKAN also consistently outperforms all methods on the cSCC and STNET datasets, highlighting its effectiveness and robust generalization across diverse ST datasets.

\subsection{Visualization of Cancer Biomarker Genes}
\label{sec:exp_visual}
To evaluate the model’s ability to capture spatial gene expression patterns both quantitatively and qualitatively, we visualize the log-normalized expression levels of two well-established cancer biomarkers in Figure~\ref{fig:visual_main}. Specifically, FN1, frequently overexpressed in breast cancer~\cite{fn1_2022_zhang}, and HSPB1, implicated in cancer progression~\cite{liang_hspb1_2023}, are highlighted. We also report their PCCs with the ground truth to assess spatial consistency and predictive accuracy. Additional visualizations are available in the supplementary materials.

\subsection{Ablation Studies}
\label{sec:exp_ablation}
To validate the effectiveness of our model design, we conducted ablation studies on several key components: the choice of text and image encoders, textual representations with different prompt strategies and LLMs, the contributions of individual modules, and the selection of fusion strategies and loss functions. For clarity, we present mean results on the STNET dataset in the main text. Consistent trends were also observed on the HER2+ and cSCC datasets, with detailed results provided in the supplementary materials.

\begin{table}[htb]
\small
  \setlength{\tabcolsep}{2pt}
  \centering
  \begin{tabular}{l|cccccc}
    \toprule
     & \multicolumn{2}{c}{Error} & \multicolumn{4}{c}{PCC} \\
    \cmidrule(lr){2-3}
    \cmidrule(lr){4-7}
    Prompt Strategy &MAE$\downarrow$&MSE$\downarrow$&ALL$\uparrow$&HPG$\uparrow$ &HEG$\uparrow$ &HVG$\uparrow$\\
    \midrule
    
    w/o text constraint&0.340&0.198&0.206&0.351&0.177&0.227\\
    w/o text summary&0.342&0.199&0.199&0.341&0.177&0.220\\
    \rowcolor{gray!20}Ours&\textbf{0.322}&\textbf{0.179}&\textbf{0.219}&\textbf{0.387}&\textbf{0.200}&\textbf{0.244}\\
    \midrule
     LLM Candidate
    &MAE$\downarrow$&MSE$\downarrow$&ALL$\uparrow$&HPG$\uparrow$ &HEG$\uparrow$ &HVG$\uparrow$\\
    \midrule
    Deepseek-R1&0.343&0.199&0.198&0.345&0.174&0.226\\
    Deepseek-v3&0.337&0.193&0.202&0.342&0.187&0.221\\
    LLaMA 2&0.339&0.194&0.193&0.337&0.165&0.218\\
    \rowcolor{gray!20}GPT-4o&\textbf{0.322}&\textbf{0.179}&\textbf{0.219}&\textbf{0.387}&\textbf{0.200}&\textbf{0.244}\\
  \bottomrule
\end{tabular}
\caption{Ablation studies on prompt strategies and LLMs.}
\label{tab:text_generation_stnet}
\end{table}

\textbf{(1) Text and Image Encoders.} For the text encoder, we evaluated four embedding models: Conch~\cite{conch} and PLIP~\cite{plip}, which are medical vision-language models, as well as BioBERT~\cite{biobert} and BioGPT~\cite{biogpt}, which are pretrained on large-scale biomedical corpora such as PubMed. For image encoders at both the region and WSI levels, we compared PLIP~\cite{plip}, Conch~\cite{conch}, UNI~\cite{uni}, ResNet50\cite{he_resnet_2016}, and ResNet18\cite{resnet18}.
As shown in Table~\ref{tab:embedding_stnet}, the performance of different text encoders was generally comparable, with BioBERT consistently achieving the best results. Among image encoders, UNI, a vision foundation model pretrained on histopathological images, achieved the highest overall performance, showing a notable improvement over other models.

\textbf{(2) Gene Semantic Representation.} 
To evaluate textual representations of genes, we compared model performance using three prompt strategies: (a) gene summaries without constraints on function or phenotype, (b) gene symbols without summaries, and (c) gene summaries enriched with specific constraints on function and phenotype. We also evaluated four widely used LLMs: DeepSeek-R1~\cite{deepseekr1}, DeepSeek-v3~\cite{liu2024deepseekv3}, LLaMA2 (7B-chat-hf)~\cite{touvron2023llama2}, and GPT-4o~\cite{openai2024gpt4o}. As shown in Table~\ref{tab:text_generation_stnet}, our proposed prompt strategy (c) achieved the best performance, as it effectively captures more informative and concise gene semantics. Among the LLMs, GPT-4o consistently outperformed the others.

\textbf{(3) Individual Modules.} We evaluated the contributions of key components in our model, including multi-scale spatial context, gene semantic features, contrastive learning, and the use of text features as Key and Value (referred to as text as KV) in dual-path contrastive learning. As shown in Table~\ref{tab:individual_modules_stnet}, removing multi-scale spatial context, gene semantics, or contrastive learning led to a decline in performance in terms of PCC. Interestingly, removing contrastive learning resulted in a slight improvement in MAE, but our configuration achieved the best overall performance across all metrics. Additionally, using text features as Query rather than as Key/Value proved more effective for multimodal integration. These findings highlight the effectiveness of our architectural design and the importance of each module.

\begin{table}[htb]
\small
  \setlength{\tabcolsep}{3pt}
  \centering
  \begin{tabular}{l|cccccc}
    \toprule 
     & \multicolumn{2}{c}{Error} & \multicolumn{4}{c}{PCC} \\
    \cmidrule(lr){2-3}
    \cmidrule(lr){4-7}
   Module &MAE$\downarrow$&MSE$\downarrow$&ALL$\uparrow$&HPG$\uparrow$ &HEG$\uparrow$ &HVG$\uparrow$\\
    \midrule
    w/o multi-scale &0.350 &0.210 &0.117&0.210&0.101&0.112\\
    w/o text &0.343&0.201&0.210&0.372&0.177&0.233\\
    w/o contrast. &\textbf{0.320}&\textbf{0.179}&0.209&0.380&0.187&0.231\\
    Text as KV&0.333&0.186&0.216&0.379&0.182&0.242\\
    \rowcolor{gray!20}Ours&0.322&\textbf{0.179}&\textbf{0.219}&\textbf{0.387}&\textbf{0.200}&\textbf{0.244}\\
  \bottomrule
\end{tabular}
\caption{Ablation studies on individual modules.}
\label{tab:individual_modules_stnet}
\end{table}

\begin{table}[htb]
\small
  \setlength{\tabcolsep}{3pt}
  \centering
  \begin{tabular}{l|cccccc}
    \toprule
     & \multicolumn{2}{c}{Error} & \multicolumn{4}{c}{PCC} \\
    \cmidrule(lr){2-3}
    \cmidrule(lr){4-7}
   Fusion Strategy &MAE$\downarrow$&MSE$\downarrow$&ALL$\uparrow$&HPG$\uparrow$ &HEG$\uparrow$ &HVG$\uparrow$\\
    \midrule
    Concat.&0.336&0.189&0.154&0.292&0.048&0.178\\
    Concat.+Trans.&0.331&0.191&0.214&0.380&0.189&0.236\\
    Sum.&0.326&\textbf{0.179}&0.151&0.282&0.053&0.171\\
    Sum.+Trans.&0.329&0.188&\textbf{0.221}&0.383&0.198&\textbf{0.247}\\
    \rowcolor{gray!20}Cross Atten.&\textbf{0.322}&\textbf{0.179}&0.219&\textbf{0.387}&\textbf{0.200}&0.244\\
    \midrule
    Loss Design    &MAE$\downarrow$&MSE$\downarrow$&ALL$\uparrow$&HPG$\uparrow$ &HEG$\uparrow$ &HVG$\uparrow$\\
    \midrule
    Fixed weights&0.338&0.191&0.148&0.276&0.053&0.168\\
    w/o distillation&0.336&0.195&0.217&0.382&\textbf{0.202}&\textbf{0.251}\\
    \rowcolor{gray!20}Ours&\textbf{0.322}&\textbf{0.179}&\textbf{0.219}&\textbf{0.387}&0.200&0.244\\
  \bottomrule
\end{tabular}
\caption{Ablation studies on fusion methods and weights.}
\label{tab:fusion_stnet}
\end{table}

\textbf{(4) Fusion Strategy and Loss Design.} 
We investigated the impact of different fusion strategies and loss function designs, as shown in Table~\ref{tab:fusion_stnet}. Specifically, we evaluated the cross-attention mechanism used in our contrastive alignment module against four alternatives: addition (Sum.), concatenation (Concat.), concatenation followed by a transformer layer (Concat. + Trans.), and addition followed by a transformer layer (Sum. + Trans.). For the loss functions, we ablated two key components, our dynamic weight balancing strategy and the knowledge distillation loss, to assess their contributions. The results show a consistent performance drop across metrics, particularly in MAE and MSE, when alternative fusion strategies or simplified loss designs are used. In contrast, our full model design achieves the best overall performance in PCC, while also maintaining the lowest MAE and MSE, demonstrating the effectiveness of both our fusion strategy and our loss function design.

\section{Conclusion}
In this study, we propose DKAN, a dual-path knowledge-augmented contrastive alignment framework that integrates high-level biological gene knowledge into multimodal feature alignment for spatial gene expression prediction. Comprehensive experiments demonstrate the superior performance of DKAN compared to existing state-of-the-art methods, highlighting the effectiveness of structured biological priors in enhancing cross-modal representation learning. This approach offers a practical pathway for linking histological morphology with spatial gene expression, supporting future discoveries in tissue microenvironments and biomarker identification.

\appendix
\setcounter{figure}{0}
\setcounter{table}{0}
\setcounter{equation}{0}
\renewcommand{\thesection}{S\arabic{section}}  
\renewcommand{\thefigure}{S\arabic{figure}} 
\renewcommand{\thetable}{S\arabic{table}}  
\renewcommand{\theequation}{S\arabic{equation}}

\section*{Supplementary Materials}
In the supplementary materials, we first present the structured prompt design used by the Large Language Model (LLM) in the gene semantic representation module. This is followed by comprehensive ablation studies on the HER2+ and cSCC datasets to assess the contributions of different components within DKAN. We also include additional ablation results on the STNET dataset, a computational comparison with baseline models, extended visualizations of cancer marker gene expression, and details on the preprocessing and evaluation protocols for gene data.

\section{LLM Prompt}
\label{sec:prompt}
\begin{figure}[htbp]
  \centering
  \includegraphics[width=\linewidth]{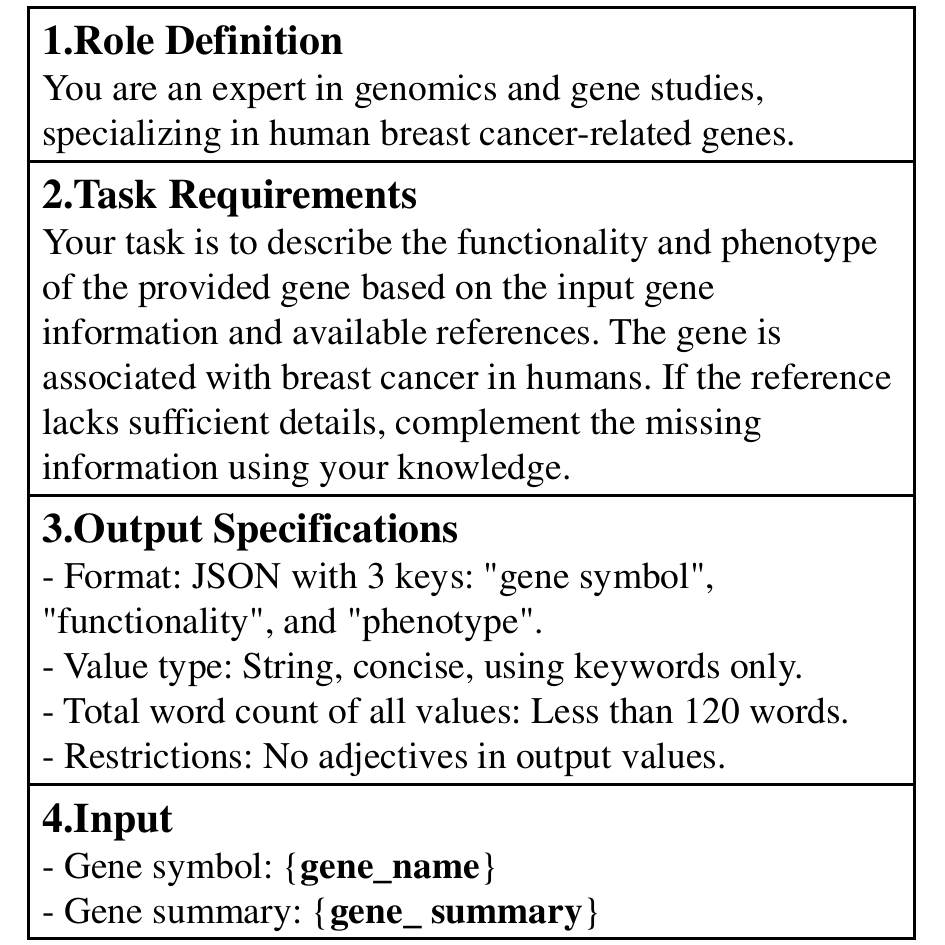}
  \caption{Prompt template for breast cancer datasets.}
  \label{fig:llm_prompt}
\end{figure}

In the gene semantic representation module, we leverage the summarization capabilities and embedded biomedical knowledge of the LLM (GPT-4o) to preserve both structural consistency and informational integrity in gene-level representations. As illustrated in Figure~\ref{fig:llm_prompt}, we designed a structured prompt tailored for breast cancer datasets, which explicitly defines the context and task requirements through three key components: role definitions, task descriptions, and output specifications. The prompt further clarifies that the provided genes are associated with human breast cancer. Gene summaries retrieved from an external database are subsequently incorporated as input to the model.

\section{Additional Ablation Studies}
\label{sec:ablation}
In this section, we present additional ablation study results on the HER2+ and cSCC datasets, shown in Table~\ref{tab:supplement_her2st} and Table~\ref{tab:supplementary_skin}, respectively. Specifically, we evaluate the effectiveness of various components, including the text encoder, image encoder, LLM prompt strategies, different LLMs, individual models, fusion strategies, and loss function designs. The evaluation is conducted using six metrics: Mean Absolute Error (MAE), Mean Squared Error (MSE), Pearson Correlation Coefficient (PCC) across all genes, as well as PCC for highly predictive genes, highly expressed genes, and highly variable genes, to comprehensively assess and validate our model design. Details of these metrics are provided in Section~\ref{sec:geneset}. Additionally, we report the complete ablation results, including standard deviations, on the STNET dataset in Table~\ref{tab:supplementary_stnet}, to supplement the findings presented in the main text. Collectively, these results provide strong empirical evidence supporting the design rationale of DKAN’s architecture.

\section{Computational Comparison}
\label{sec:computation}
To assess the computational complexity of our proposed method relative to existing State-Of-The-Art (SOTA) models, we report the number of FLOPs, model parameters, training time, and inference time in Table~\ref{tab:computation} on cSCC dataset. All experiments were conducted on an NVIDIA A800 GPU to ensure fair and consistent comparisons. Inference time was measured as the average per spot across all cross-validation folds. The results demonstrate that the integration of gene semantic representation and dual-path contrastive learning introduces minimal computational overhead. Moreover, DKAN exhibits strong computational efficiency, particularly in terms of training and inference time, when compared to baseline methods.

\begin{table*}
  \setlength{\tabcolsep}{3pt}
  \centering
  \begin{tabular}{l|cccccc}
    \toprule
     & \multicolumn{2}{c}{Error} & \multicolumn{4}{c}{PCC} \\
    \cmidrule(lr){2-3}
    \cmidrule(lr){4-7}
    Text Emb. &MAE$\downarrow$&MSE$\downarrow$&ALL$\uparrow$&HPG$\uparrow$ &HEG$\uparrow$ &HVG$\uparrow$\\
    \midrule
    Conch&$0.363\pm0.04$&$0.227\pm0.06$&$\textbf{0.333}\pm\textbf{0.13}$&$0.528\pm0.15$&$0.312\pm0.09$&$\textbf{0.305}\pm\textbf{0.06}$\\
    PLIP&$0.372\pm0.05$&$0.236\pm0.07$&$0.324\pm0.14$&$0.525\pm0.16$&$0.310\pm0.10$&$0.299\pm0.08$\\
    BioGPT&$0.377\pm0.07$&$0.243\pm0.09$&$0.329\pm0.14$&$0.522\pm0.15$&$0.313\pm0.10$&$0.302\pm0.07$\\
    \rowcolor{gray!20}BioBERT&$\textbf{0.361}\pm\textbf{0.04}$&$\textbf{0.224}\pm\textbf{0.06}$&$0.330\pm0.13$&$\textbf{0.531}\pm\textbf{0.15}$&$\textbf{0.317}\pm\textbf{0.09}$&$0.304\pm0.07$\\
    \midrule
    Image Emb. &MAE$\downarrow$&MSE$\downarrow$&ALL$\uparrow$&HPG$\uparrow$ &HEG$\uparrow$ &HVG$\uparrow$\\
    \midrule
    ResNet18&$0.368\pm0.04$&$0.233\pm0.05$&$0.312\pm0.14$&$0.500\pm0.17$&$0.280\pm0.10$&$0.272\pm0.07$\\
    ResNet50&$0.377\pm0.05$&$0.244\pm0.07$&$0.303\pm0.13$&$0.489\pm0.14$&$0.283\pm0.09$&$0.269\pm0.05$\\
    PLIP&$0.398\pm0.06$&$0.266\pm0.07$&$0.298\pm0.14$&$0.483\pm0.17$&$0.267\pm0.10$&$0.249\pm0.07$\\
    Conch&$\textbf{0.360}\pm\textbf{0.05}$&$\textbf{0.222}\pm\textbf{0.06}$&$0.326\pm0.13$&$0.519\pm0.15$&$0.301\pm0.10$&$0.294\pm0.06$\\
    \rowcolor{gray!20}UNI&$0.361\pm0.04$&$0.224\pm0.06$&$\textbf{0.330}\pm\textbf{0.13}$&$\textbf{0.531}\pm\textbf{0.15}$&$\textbf{0.317}\pm\textbf{0.09}$&$\textbf{0.304}\pm\textbf{0.07}$\\
    \midrule
    Prompt Strategy &MAE$\downarrow$&MSE$\downarrow$&ALL$\uparrow$&HPG$\uparrow$ &HEG$\uparrow$ &HVG$\uparrow$\\
    \midrule
    
    w/o text constraint&$0.367\pm0.05$&$0.231\pm0.05$&$0.310\pm0.14$&$0.492\pm0.17$&$0.277\pm0.09$&$0.256\pm0.06$\\
    w/o text summary&$0.368\pm0.04$&$0.232\pm0.05$&$0.308\pm0.14$&$0.488\pm0.17$&$0.273\pm0.10$&$0.259\pm0.06$\\
    \rowcolor{gray!20}Ours&$\textbf{0.361}\pm\textbf{0.04}$&$\textbf{0.224}\pm\textbf{0.06}$&$\textbf{0.330}\pm\textbf{0.13}$&$\textbf{0.531}\pm\textbf{0.15}$&$\textbf{0.317}\pm\textbf{0.09}$&$\textbf{0.304}\pm\textbf{0.07}$\\
    \midrule
     LLM Candidate &MAE$\downarrow$&MSE$\downarrow$&ALL$\uparrow$&HPG$\uparrow$ &HEG$\uparrow$ &HVG$\uparrow$\\
    \midrule
    Deepseek-R1&$0.371\pm0.04$&$0.237\pm0.05$&$0.302\pm0.14$&$0.490\pm0.17$&$0.270\pm0.09$&$0.255\pm0.06$\\
    Deepseek-v3&$0.370\pm0.04$&$0.236\pm0.06$&$0.308\pm0.14$&$0.497\pm0.16$&$0.283\pm0.10$&$0.266\pm0.06$\\
    LLaMA 2 &$\textbf{0.358}\pm\textbf{0.05}$&$\textbf{0.221}\pm\textbf{0.05}$&$0.307\pm0.14$&$0.487\pm0.17$&$0.280\pm0.09$&$0.267\pm0.06$\\
    \rowcolor{gray!20}GPT-4o&$0.361\pm0.04$&$0.224\pm0.06$&$\textbf{0.330}\pm\textbf{0.13}$&$\textbf{0.531}\pm\textbf{0.15}$&$\textbf{0.317}\pm\textbf{0.09}$&$\textbf{0.304}\pm\textbf{0.07}$\\
    \midrule
    Module &MAE$\downarrow$&MSE$\downarrow$&ALL$\uparrow$&HPG$\uparrow$ &HEG$\uparrow$ &HVG$\uparrow$\\
    \midrule
    w/o multi-scale 
    &$0.390\pm0.02$ &$0.253\pm0.03$ &$0.210\pm0.13$ &$0.336\pm0.17$ &$0.166\pm0.09$ &$0.144\pm0.06$\\ 
    
    w/o text &$0.371\pm0.04$&$0.241\pm0.06$&$0.311\pm0.13$&$0.512\pm0.16$&$0.288\pm0.09$&$0.287\pm0.06$\\
    w/o contrast &$0.369\pm0.05$&$0.234\pm0.06$&$0.303\pm0.15$&$0.492\pm0.17$&$0.277\pm0.11$&$0.275\pm0.07$\\

    Text as KV&$0.372\pm0.04$&$0.242\pm0.05$&$\textbf{0.330}\pm\textbf{0.14}$&$0.527\pm0.15$&$0.307\pm0.10$&$0.303\pm0.07$\\
    \rowcolor{gray!20}Ours&$\textbf{0.361}\pm\textbf{0.04}$&$\textbf{0.224}\pm\textbf{0.06}$&$\textbf{0.330}\pm\textbf{0.13}$&$\textbf{0.531}\pm\textbf{0.15}$&$\textbf{0.317}\pm\textbf{0.09}$&$\textbf{0.304}\pm\textbf{0.07}$\\
    \midrule
    Fusion Strategy &MAE$\downarrow$&MSE$\downarrow$&ALL$\uparrow$&HPG$\uparrow$ &HEG$\uparrow$ &HVG$\uparrow$\\
    \midrule
    Concat.&$0.380\pm0.07$&$0.246\pm0.09$&$0.254\pm0.18$&$0.471\pm0.22$&$0.142\pm0.16$&$0.105\pm0.05$\\
    Concat. + Trans.&$0.370\pm0.06$&$0.241\pm0.07$&$\textbf{0.331}\pm\textbf{0.14}$&$0.530\pm0.14$&$\textbf{0.317}\pm\textbf{0.10}$&$0.304\pm0.07$\\
    Sum.&$0.370\pm0.06$&$0.231\pm0.08$&$0.259\pm0.18$&$0.494\pm0.19$&$0.147\pm0.16$&$0.108\pm0.14$\\
    Sum. + Trans.&$0.363\pm0.06$&$0.235\pm0.07$&$0.329\pm0.13$&$0.524\pm0.16$&$0.307\pm0.09$&$\textbf{0.312}\pm\textbf{0.06}$\\
    \rowcolor{gray!20}Cross Atten.&$\textbf{0.361}\pm\textbf{0.04}$&$\textbf{0.224}\pm\textbf{0.06}$&$0.330\pm0.13$&$\textbf{0.531}\pm\textbf{0.15}$&$\textbf{0.317}\pm\textbf{0.09}$&$0.304\pm0.07$\\
    \midrule

    Loss Design    &MAE$\downarrow$&MSE$\downarrow$&ALL$\uparrow$&HPG$\uparrow$ &HEG$\uparrow$ &HVG$\uparrow$\\
    \midrule
    Fixed weights&$0.385\pm0.07$&$0.250\pm0.10$&$0.258\pm0.17$&$0.470\pm0.20$&$0.147\pm0.16$&$0.108\pm0.13$\\
    w/o distillation&$0.392\pm0.04$&$0.256\pm0.06$&$0.316\pm0.13$&$0.517\pm0.16$&$0.298\pm0.09$&$0.300\pm0.07$\\
    \rowcolor{gray!20}Ours&$\textbf{0.361}\pm\textbf{0.04}$&$\textbf{0.224}\pm\textbf{0.06}$&$\textbf{0.330}\pm\textbf{0.13}$&$\textbf{0.531}\pm\textbf{0.15}$&$\textbf{0.317}\pm\textbf{0.09}$&$\textbf{0.304}\pm\textbf{0.07}$\\
  \bottomrule
\end{tabular}
\caption{Additional ablation studies on HER2+ dataset.}
\label{tab:supplement_her2st}
\end{table*}

\begin{table*}
  \setlength{\tabcolsep}{3pt}
  \centering
  \begin{tabular}{l|cccccc}
    \toprule
     & \multicolumn{2}{c}{Error} & \multicolumn{4}{c}{PCC} \\
    \cmidrule(lr){2-3}
    \cmidrule(lr){4-7}
    Text Emb. &MAE$\downarrow$&MSE$\downarrow$&ALL$\uparrow$&HPG$\uparrow$ &HEG$\uparrow$ &HVG$\uparrow$\\
    \midrule
    Conch&$0.402\pm0.05$&$0.263\pm0.07$&$0.410\pm0.07$&$0.504\pm0.08$&$\textbf{0.352}\pm\textbf{0.08}$&$\textbf{0.335}\pm\textbf{0.08}$\\
    PLIP&$0.395\pm0.05$&$0.254\pm0.07$&$\textbf{0.413}\pm\textbf{0.08}$&$0.516\pm0.07$&$\textbf{0.352}\pm\textbf{0.08}$&$0.327\pm0.08$\\
    BioGPT&$0.400\pm0.06$&$0.259\pm0.08$&$0.412\pm0.08$&$\textbf{0.517}\pm\textbf{0.07}$&$0.341\pm0.09$&$0.311\pm0.08$\\
    \rowcolor{gray!20}BioBERT&$\textbf{0.383}\pm\textbf{0.05}$&$\textbf{0.239}\pm0.06$&$0.407\pm0.08$&$0.508\pm0.08$&$0.346\pm0.09$&$0.321\pm0.08$\\
    \midrule
    Image Emb. &MAE$\downarrow$&MSE$\downarrow$&ALL$\uparrow$&HPG$\uparrow$ &HEG$\uparrow$ &HVG$\uparrow$\\
    \midrule
    ResNet18&$0.404\pm0.05$&$0.263\pm0.06$&$0.390\pm0.07$&$0.496\pm0.06$&$0.308\pm0.08$&$0.289\pm0.08$\\
    ResNet50&$0.408\pm0.04$&$0.267\pm0.04$&$0.381\pm0.06$&$0.484\pm0.06$&$0.294\pm0.06$&$0.289\pm0.06$\\
    PLIP&$0.392\pm0.03$&$0.245\pm0.04$&$0.395\pm0.07$&$0.502\pm0.06$&$0.320\pm0.08$&$0.292\pm0.08$\\
    Conch&$0.386\pm0.05$&$0.241\pm0.05$&$0.403\pm0.07$&$0.510\pm0.07$&$0.339\pm0.08$&$0.309\pm0.08$\\
    \rowcolor{gray!20}UNI&$\textbf{0.383}\pm\textbf{0.05}$&$\textbf{0.239}\pm\textbf{0.06}$&$\textbf{0.407}\pm\textbf{0.08}$&$\textbf{0.508}\pm\textbf{0.08}$&$\textbf{0.346}\pm\textbf{0.09}$&$\textbf{0.321}\pm\textbf{0.08}$\\
    \midrule
    Prompt Strategy &MAE$\downarrow$&MSE$\downarrow$&ALL$\uparrow$&HPG$\uparrow$ &HEG$\uparrow$ &HVG$\uparrow$\\
    \midrule
    
    w/o text constraint&$0.403\pm0.05$&$0.261\pm0.06$&$0.394\pm0.07$&$0.497\pm0.06$&$0.320\pm0.08$&$0.300\pm0.08$\\
    w/o text summary&$0.412\pm0.06$&$0.271\pm0.07$&$0.387\pm0.07$&$0.491\pm0.07$&$0.315\pm0.09$&$0.290\pm0.08$\\
    \rowcolor{gray!20}Ours&$\textbf{0.383}\pm\textbf{0.05}$&$\textbf{0.239}\pm\textbf{0.06}$&$\textbf{0.407}\pm\textbf{0.08}$&$\textbf{0.508}\pm\textbf{0.08}$&$\textbf{0.346}\pm\textbf{0.09}$&$\textbf{0.321}\pm\textbf{0.08}$\\
    \midrule
     LLM Candidate
    &MAE$\downarrow$&MSE$\downarrow$&ALL$\uparrow$&HPG$\uparrow$ &HEG$\uparrow$ &HVG$\uparrow$\\
    \midrule
    Deepseek-R1&$0.407\pm0.06$&$0.267\pm0.07$&$0.391\pm0.07$&$0.502\pm0.06$&$0.312\pm0.07$&$0.288\pm0.06$\\
    Deepseek-V3&$0.405\pm0.06$&$0.265\pm0.08$&$0.391\pm0.07$&$0.496\pm0.06$&$0.317\pm0.08$&$0.298\pm0.07$\\
    Llama2&$0.407\pm0.05$&$0.263\pm0.06$&$0.386\pm0.07$&$0.491\pm0.06$&$0.314\pm0.07$&$0.295\pm0.07$\\
    \rowcolor{gray!20}GPT-4o&$\textbf{0.383}\pm\textbf{0.05}$&$\textbf{0.239}\pm\textbf{0.06}$&$\textbf{0.407}\pm\textbf{0.08}$&$\textbf{0.508}\pm\textbf{0.08}$&$\textbf{0.346}\pm\textbf{0.09}$&$\textbf{0.321}\pm\textbf{0.08}$\\

    \midrule
    Module &MAE$\downarrow$&MSE$\downarrow$&ALL$\uparrow$&HPG$\uparrow$ &HEG$\uparrow$ &HVG$\uparrow$\\
    \midrule
    w/o multi-scale
    &$0.425\pm0.04$ &$0.281\pm0.04$ &$0.300\pm0.07$ &$0.402\pm0.06$ &$0.229\pm0.07$ &$0.205\pm0.07$\\
    w/o text &$0.410\pm0.07$&$0.278\pm0.09$&$0.326\pm0.05$&$0.447\pm0.06$&$0.258\pm0.07$&$0.265\pm0.07$\\
    w/o contrast &$0.400\pm0.04$&$0.253\pm0.05$&$0.388\pm0.09$&$0.492\pm0.07$&$0.318\pm0.11$&$0.289\pm0.10$\\
    Text as KV&$0.395\pm0.05$&$0.254\pm0.06$&$0.368\pm0.06$&$0.486\pm0.06$&$0.275\pm0.09$&$0.277\pm0.09$\\
    \rowcolor{gray!20}Ours&$\textbf{0.383}\pm\textbf{0.05}$&$\textbf{0.239}\pm\textbf{0.06}$&$\textbf{0.407}\pm\textbf{0.08}$&$\textbf{0.508}\pm\textbf{0.08}$&$\textbf{0.346}\pm\textbf{0.09}$&$\textbf{0.321}\pm\textbf{0.08}$\\

    \midrule
   Fusion Strategy &MAE$\downarrow$&MSE$\downarrow$&ALL$\uparrow$&HPG$\uparrow$ &HEG$\uparrow$ &HVG$\uparrow$\\
    \midrule
    Concat.&$0.410\pm0.08$&$0.271\pm0.10$&$0.356\pm0.08$&$0.498\pm0.06$&$0.230\pm0.12$&$0.175\pm0.11$\\
    Concat. + Trans.&$0.390\pm0.07$&$0.251\pm0.08$&$0.388\pm0.08$&$0.498\pm0.07$&$0.302\pm0.08$&$0.293\pm0.09$\\
    Sum.&$0.402\pm0.06$&$0.257\pm0.08$&$0.352\pm0.08$&$0.491\pm0.07$&$0.229\pm0.12$&$0.179\pm0.11$\\
    Sum. + Trans.&$0.399\pm0.06$&$0.260\pm0.07$&$0.400\pm0.08$&$0.498\pm0.08$&$0.331\pm0.09$&$0.324\pm0.09$\\
    \rowcolor{gray!20}Cross Atten.&$\textbf{0.383}\pm\textbf{0.05}$&$\textbf{0.239}\pm\textbf{0.06}$&$\textbf{0.407}\pm\textbf{0.08}$&$\textbf{0.508}\pm\textbf{0.08}$&$\textbf{0.346}\pm\textbf{0.09}$&$\textbf{0.321}\pm\textbf{0.08}$\\
    \midrule
    Loss Design    &MAE$\downarrow$&MSE$\downarrow$&ALL$\uparrow$&HPG$\uparrow$ &HEG$\uparrow$ &HVG$\uparrow$\\
    \midrule
    Fixed weights&$0.408\pm0.08$&$0.266\pm0.09$&$0.338\pm0.09$&$0.479\pm0.07$&$0.220\pm0.12$&$0.169\pm0.11$\\
    w/o distillation&$0.391\pm0.03$&$0.248\pm0.03$&$0.402\pm0.07$&$0.503\pm0.07$&$0.337\pm0.09$&$0.323\pm0.10$\\
    \rowcolor{gray!20}Ours&$\textbf{0.383}\pm\textbf{0.05}$&$\textbf{0.239}\pm\textbf{0.06}$&$\textbf{0.407}\pm\textbf{0.08}$&$\textbf{0.508}\pm\textbf{0.08}$&$\textbf{0.346}\pm\textbf{0.09}$&$\textbf{0.321}\pm\textbf{0.08}$\\
  \bottomrule
\end{tabular}
\caption{Additional ablation studies on cSCC dataset.}
\label{tab:supplementary_skin}
\end{table*}

\begin{table*}
  \setlength{\tabcolsep}{3pt}
  \centering
  \begin{tabular}{l|cccccc}
    \toprule
     & \multicolumn{2}{c}{Error} & \multicolumn{4}{c}{PCC} \\
    \cmidrule(lr){2-3}
    \cmidrule(lr){4-7}
    Text Emb. &MAE$\downarrow$&MSE$\downarrow$&ALL$\uparrow$&HPG$\uparrow$ &HEG$\uparrow$ &HVG$\uparrow$\\
    \midrule
    Conch&$0.324\pm0.02$&$0.181\pm0.02$&$0.211\pm0.06$&$0.374\pm0.09$&$0.194\pm0.06$&$0.230\pm0.07$\\
    PLIP&$0.323\pm0.02$&$\textbf{0.179}\pm\textbf{0.02}$&$0.217\pm0.07$&$0.381\pm0.09$&$0.194\pm0.06$&$\textbf{0.244}\pm\textbf{0.06}$\\
    BioGPT&$0.327\pm0.02$&$0.183\pm0.02$&$0.217\pm0.07$&$0.385\pm0.09$&$0.193\pm0.06$&$0.239\pm0.06$\\
    \rowcolor{gray!20}BioBERT&$\textbf{0.322}\pm\textbf{0.02}$&$\textbf{0.179}\pm\textbf{0.02}$&$\textbf{0.219}\pm\textbf{0.07}$&$\textbf{0.387}\pm\textbf{0.09}$&$\textbf{0.200}\pm\textbf{0.06}$&$\textbf{0.244}\pm\textbf{0.07}$\\
    \midrule
    Image Emb. &MAE$\downarrow$&MSE$\downarrow$&ALL$\uparrow$&HPG$\uparrow$ &HEG$\uparrow$ &HVG$\uparrow$\\
    \midrule
    ResNet18&$0.341\pm0.01$&$0.197\pm0.02$&$0.202\pm0.06$&$0.347\pm0.09$&$0.176\pm0.06$&$0.224\pm0.06$\\
    ResNet50&$0.354\pm0.01$&$0.212\pm0.02$&$0.196\pm0.06$&$0.349\pm0.10$&$0.167\pm0.06$&$0.215\pm0.07$\\
    PLIP&$0.333\pm0.02$&$0.188\pm0.02$&$0.190\pm0.07$&$0.333\pm0.11$&$0.157\pm0.06$&$0.211\pm0.07$\\
    Conch&$0.324\pm0.02$&$\textbf{0.177}\pm\textbf{0.02}$&$0.209\pm0.07$&$0.369\pm0.10$&$0.179\pm0.06$&$0.231\pm0.07$\\
    \rowcolor{gray!20}UNI&$\textbf{0.322}\pm\textbf{0.02}$&$0.179\pm0.02$&$\textbf{0.219}\pm\textbf{0.07}$&$\textbf{0.387}\pm\textbf{0.09}$&$\textbf{0.200}\pm\textbf{0.06}$&$\textbf{0.244}\pm\textbf{0.07}$\\
 \midrule
    Prompt Strategy &MAE$\downarrow$&MSE$\downarrow$&ALL$\uparrow$&HPG$\uparrow$ &HEG$\uparrow$ &HVG$\uparrow$\\
    \midrule
    w/o text constraint&$0.340\pm0.02$ &$0.198\pm0.02$ &$0.206\pm0.06$ &$0.351\pm0.09$ &$0.177\pm0.06$ &$0.227\pm0.06$ \\
    w/o text summary&$0.342\pm0.02$ &$0.199\pm0.03$ &$0.199\pm0.06$ &$0.341\pm0.10$ &$0.177\pm0.06$ &$0.220\pm0.07$ \\
    \rowcolor{gray!20} Ours&$\textbf{0.322}\pm\textbf{0.02}$ &$\textbf{0.179}\pm\textbf{0.02}$ &$\textbf{0.219}\pm\textbf{0.07}$ &$\textbf{0.387}\pm\textbf{0.09}$ &$\textbf{0.200}\pm\textbf{0.06}$ &$\textbf{0.244}\pm\textbf{0.07}$\\
    \midrule
    LLM Candidate &MAE$\downarrow$ &MSE$\downarrow$ &ALL$\uparrow$ &HPG$\uparrow$ &HEG$\uparrow$ &HVG$\uparrow$\\
    \midrule
    Deepseek-R1&$0.343\pm0.02$ &$0.199\pm0.02$ &$0.198\pm0.06$ &$0.345\pm0.09$ &$0.174\pm0.05$ &$0.226\pm0.07$\\
    Deepseek-v3&$0.337\pm0.02$&$0.193\pm0.02$&$0.202\pm0.07$&$0.342\pm0.09$&$0.187\pm0.06$&$0.221\pm0.07$\\
    LLaMA 2&$0.339\pm0.01$&$0.194\pm0.02$&$0.193\pm0.07$&$0.337\pm0.10$&$0.165\pm0.07$&$0.218\pm0.06$\\
    \rowcolor{gray!20} GPT-4o &$\textbf{0.322}\pm\textbf{0.02}$ &$\textbf{0.179}\pm\textbf{0.02}$ &$\textbf{0.219}\pm\textbf{0.07}$ &$\textbf{0.387}\pm\textbf{0.09}$ &$\textbf{0.200}\pm\textbf{0.06}$ &$\textbf{0.244}\pm\textbf{0.07}$\\
     \midrule
    Module &MAE$\downarrow$&MSE$\downarrow$&ALL$\uparrow$&HPG$\uparrow$ &HEG$\uparrow$ &HVG$\uparrow$\\
    \midrule
    w/o multi-scale
    &$0.350\pm0.02$ &$0.210\pm0.02$ &$0.117\pm0.05$ &$0.210\pm0.09$ &$0.101\pm0.05$ &$0.112\pm0.04$\\ 
    w/o text &$0.343\pm0.02$&$0.201\pm0.03$&$0.210\pm0.07$&$0.372\pm0.09$&$0.177\pm0.06$&$0.233\pm0.06$\\
    w/o contrast &$\textbf{0.320}\pm\textbf{0.02}$&$\textbf{0.179}\pm\textbf{0.02}$&$0.209\pm0.07$&$0.380\pm0.09$&$0.187\pm0.06$&$0.231\pm0.06$\\
    Text as KV&$0.333\pm0.02$&$0.186\pm0.02$&$0.216\pm0.06$&$0.379\pm0.09$&$0.182\pm0.06$&$0.242\pm0.06$\\
    \rowcolor{gray!20}Ours&$0.322\pm0.02$&$\textbf{0.179}\pm\textbf{0.02}$&$\textbf{0.219}\pm\textbf{0.07}$&$\textbf{0.387}\pm\textbf{0.09}$&$\textbf{0.200}\pm\textbf{0.06}$&$\textbf{0.244}\pm\textbf{0.07}$\\
    \midrule
   Fusion Strategy &MAE$\downarrow$&MSE$\downarrow$&ALL$\uparrow$&HPG$\uparrow$ &HEG$\uparrow$ &HVG$\uparrow$\\
    \midrule
    Concat.&$0.336\pm0.02$&$0.189\pm0.03$&$0.154\pm0.07$&$0.292\pm0.13$&$0.048\pm0.06$&$0.178\pm0.08$\\
    Concat. + Trans.&$0.331\pm0.02$&$0.191\pm0.03$&$0.214\pm0.06$&$0.380\pm0.08$&$0.189\pm0.06$&$0.236\pm0.06$\\
    Sum.&$0.326\pm0.02$&$\textbf{0.179}\pm\textbf{0.03}$&$0.151\pm0.07$&$0.282\pm0.14$&$0.053\pm0.05$&$0.171\pm0.08$\\
    Sum. + Trans.&$0.329\pm0.03$&$0.188\pm0.03$&$\textbf{0.221}\pm\textbf{0.07}$&$0.383\pm0.09$&$0.198\pm0.06$&$\textbf{0.247}\pm\textbf{0.07}$\\
    \rowcolor{gray!20}Cross Atten.&$\textbf{0.322}\pm\textbf{0.02}$&$\textbf{0.179}\pm\textbf{0.02}$&$0.219\pm0.07$&$\textbf{0.387}\pm\textbf{0.09}$&$\textbf{0.200}\pm\textbf{0.06}$&$0.244\pm0.07$\\
    \midrule
    Loss Design    &MAE$\downarrow$&MSE$\downarrow$&ALL$\uparrow$&HPG$\uparrow$ &HEG$\uparrow$ &HVG$\uparrow$\\
    \midrule
    Fixed weights&$0.338\pm0.03$&$0.191\pm0.03$&$0.148\pm0.07$&$0.276\pm0.14$&$0.053\pm0.05$&$0.168\pm0.07$\\
    w/o distillation&$0.336\pm0.02$&$0.195\pm0.02$&$0.217\pm0.07$&$0.382\pm0.10$&$\textbf{0.202}\pm\textbf{0.06}$&$\textbf{0.251}\pm\textbf{0.06}$\\
    \rowcolor{gray!20}Ours&$\textbf{0.322}\pm\textbf{0.02}$&$\textbf{0.179}\pm\textbf{0.02}$&$\textbf{0.219}\pm\textbf{0.07}$&$\textbf{0.387}\pm\textbf{0.09}$&$0.200\pm0.06$&$0.244\pm0.07$\\
  \bottomrule
\end{tabular}
\caption{Ablation studies on STNET dataset with standard deviations to supplement results in the main text.  }
\label{tab:supplementary_stnet}
\end{table*}

\begin{table}[htbp]
\centering
\begin{adjustbox}{width=0.5\textwidth}
\begin{tabular}{l|rrrr}
\toprule
 Method & FLOPs (G) & \# Parameter (M) & Training Time (h) & Inference Time (s) \\ \midrule
ST-Net &2.865 &7.210 &4.119 &0.491\\
Hist2ST &375.296 &778.240 &7.199 &0.575 \\
THItoGene &79.811 &63.096 &6.024 &0.416 \\
HisToGene &52.605 &227.328 &40.725 &0.158 \\
M2OST &0.165 &208.896 & 59.263 &0.165
 \\
EGN &5.479 & 133.120 & 44.834 &0.107\\
BLEEP &2940.416  &24.229  &4.732& 0.461 \\
mclSTExp &2.866 &41.459  &20.413 & 0.587\\
TRIPLEX &4.167	&27.460  &4.758  & 0.136\\
\midrule
DKAN w/o text &12.260 &77.288 &3.500 &0.107\\
DKAN w/o dual path &15.511 &73.649 &15.183 &0.102\\
DKAN & 15.838	& 87.172	&3.711  & 0.158\\
\bottomrule
\end{tabular}
\end{adjustbox}
\caption{\label{tab:computation} Computational comparisons on cSCC dataset.}
\end{table}

\begin{figure}[htbp]
  \centering
  \includegraphics[width=\linewidth]{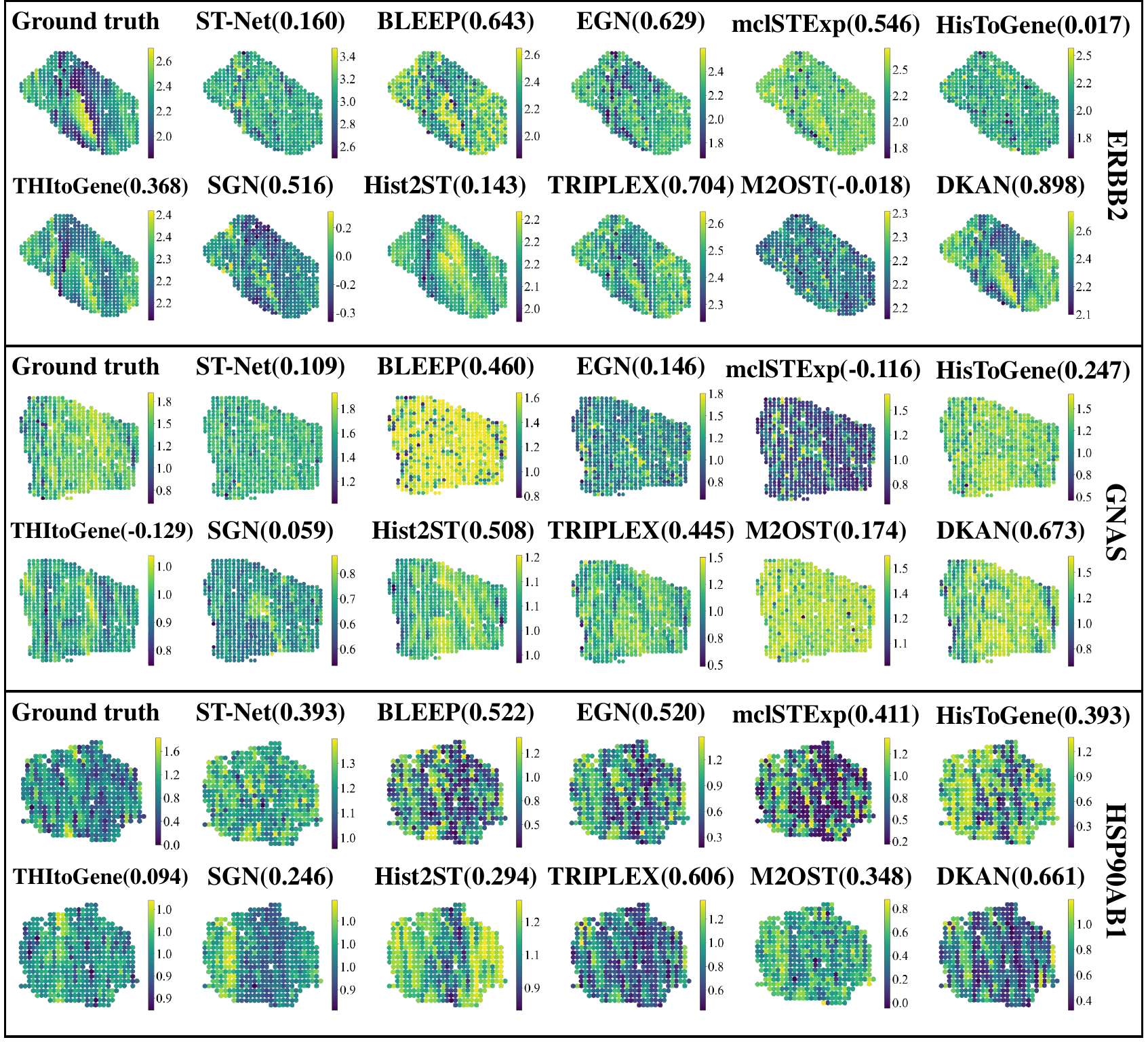}
  \caption{Visualization of Cancer Marker Gene Expression on HER2+ dataset.}
  \label{fig:visual_her2st}
\end{figure}

\begin{figure}[htbp]
  \centering
  \includegraphics[width=\linewidth]{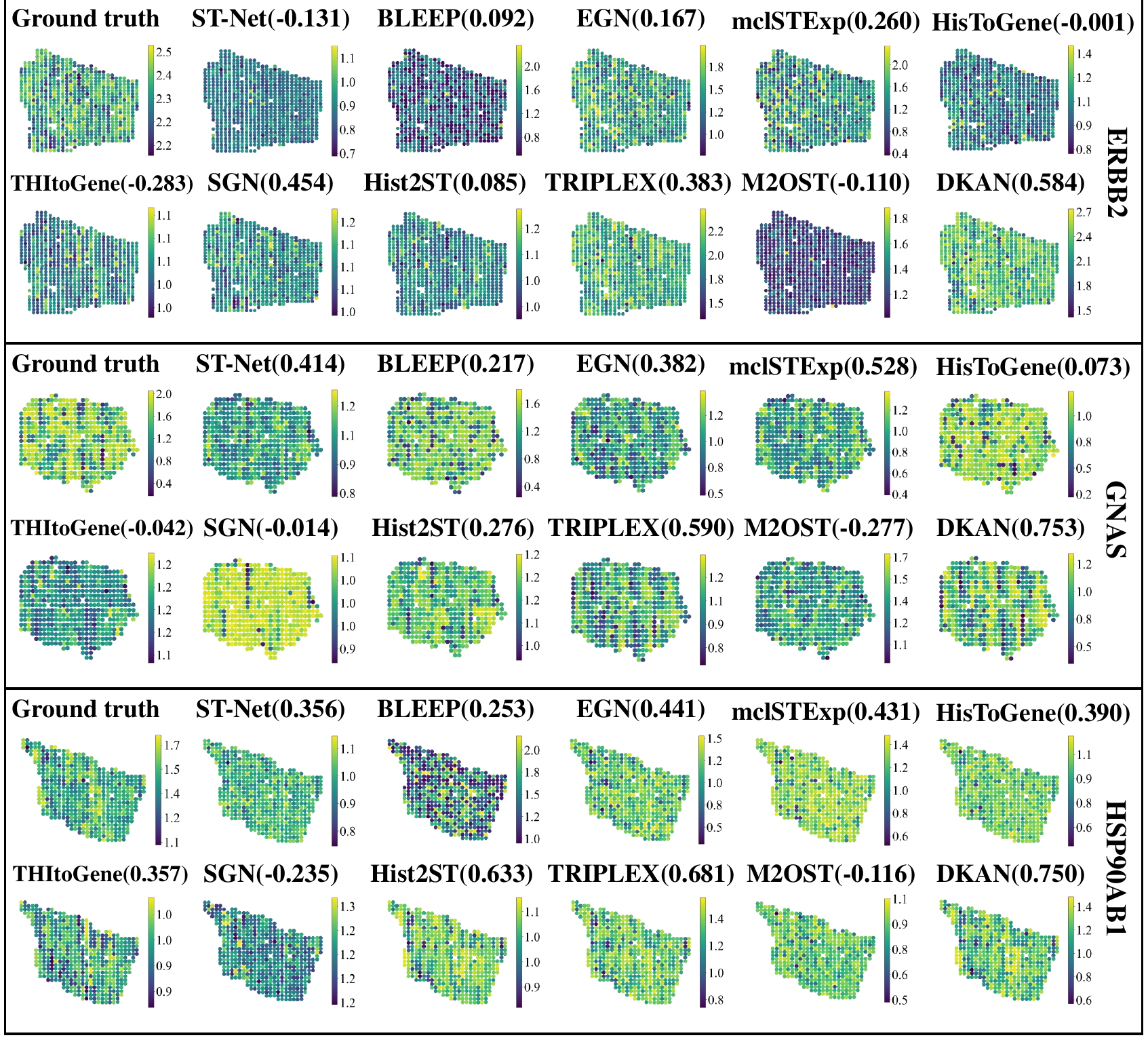}
  \caption{Visualization of Cancer Marker Gene Expression on STNET dataset.}
  \label{fig:visual_stnet}
\end{figure}

\section{Additional Visualization Results}
\label{sec:visual}
\begin{figure}[htbp]
  \centering
  \includegraphics[width=\linewidth]{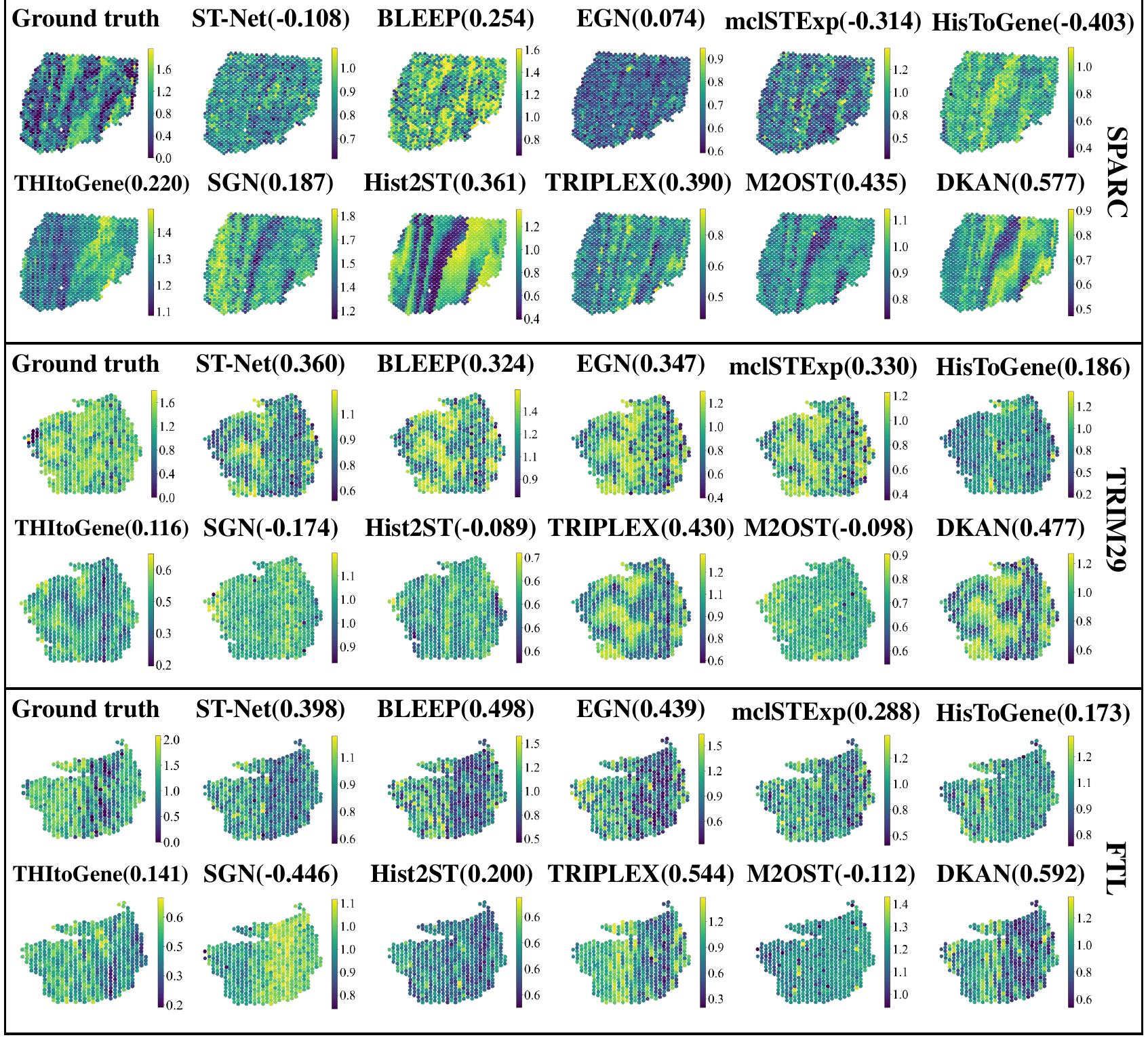}
  \caption{Visualization of Cancer Marker Gene Expression on cSCC dataset.}
  \label{fig:visual_skin}
\end{figure}

In this section, we present additional visualizations of cancer marker gene expression across the three datasets, along with the corresponding PCC values in comparison to SOTA baselines. We select genes that exhibit strong correlations with specific cancer types in their respective datasets. Specifically, ERBB2, GNAS, and HSP90AB1 are selected for the HER2+ and STNET datasets (Figures~\ref{fig:visual_her2st} and~\ref{fig:visual_stnet}), while SPARC, TRIM29, and FTL are chosen for the cSCC dataset (Figure~\ref{fig:visual_skin}). These visualizations further illustrate the effectiveness of our model in capturing both the absolute expression levels and expression trends of key marker genes, underscoring DKAN’s superior performance in gene expression prediction.

\begin{figure}[htbp]
  \centering
  \includegraphics[width=0.93\linewidth]{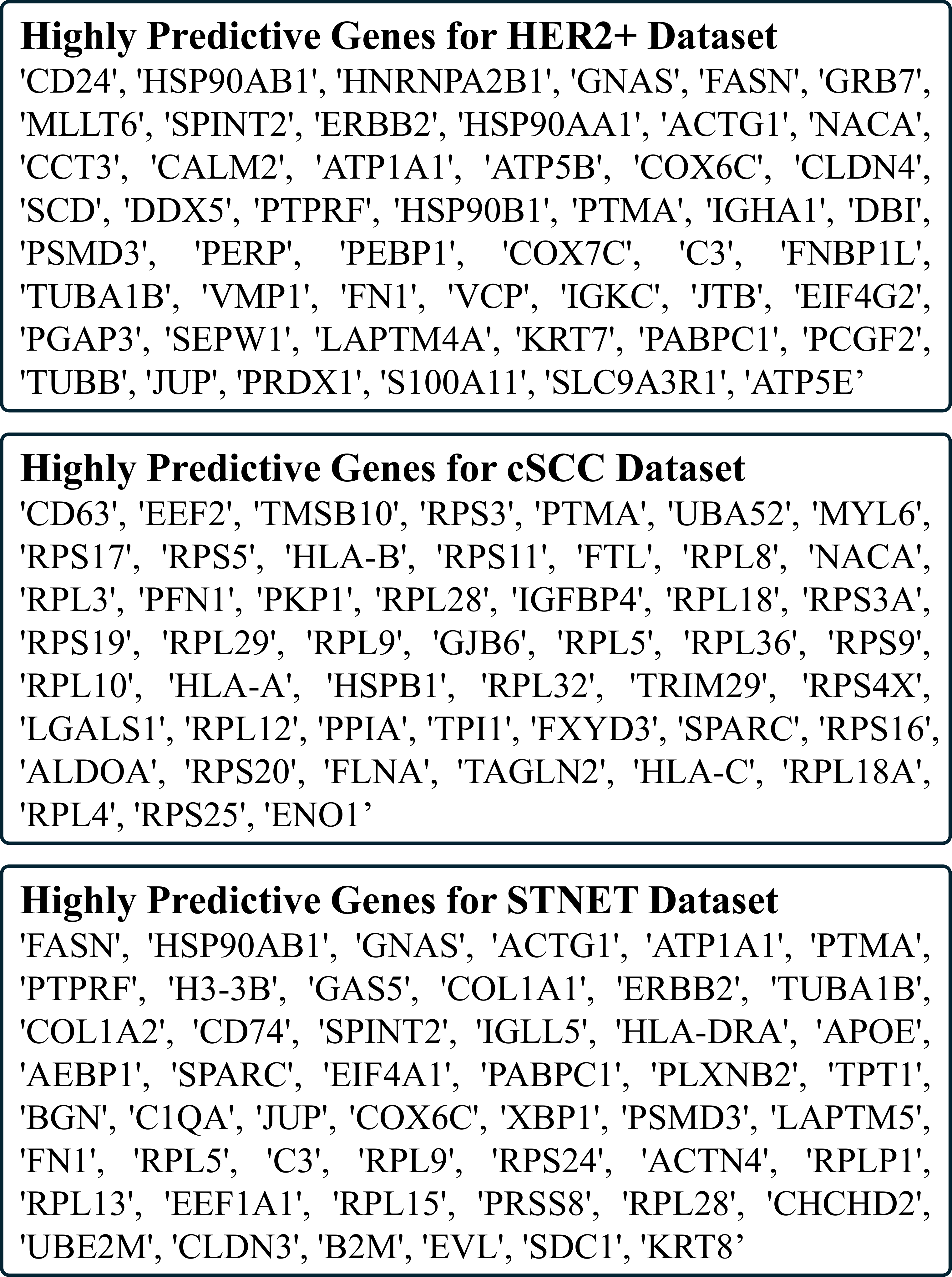}
  \caption{Highly predictive genes for all datasets.}
  \label{fig:hpg}
\end{figure}

\begin{figure}[htbp]
  \centering
  \includegraphics[width=0.93\linewidth]{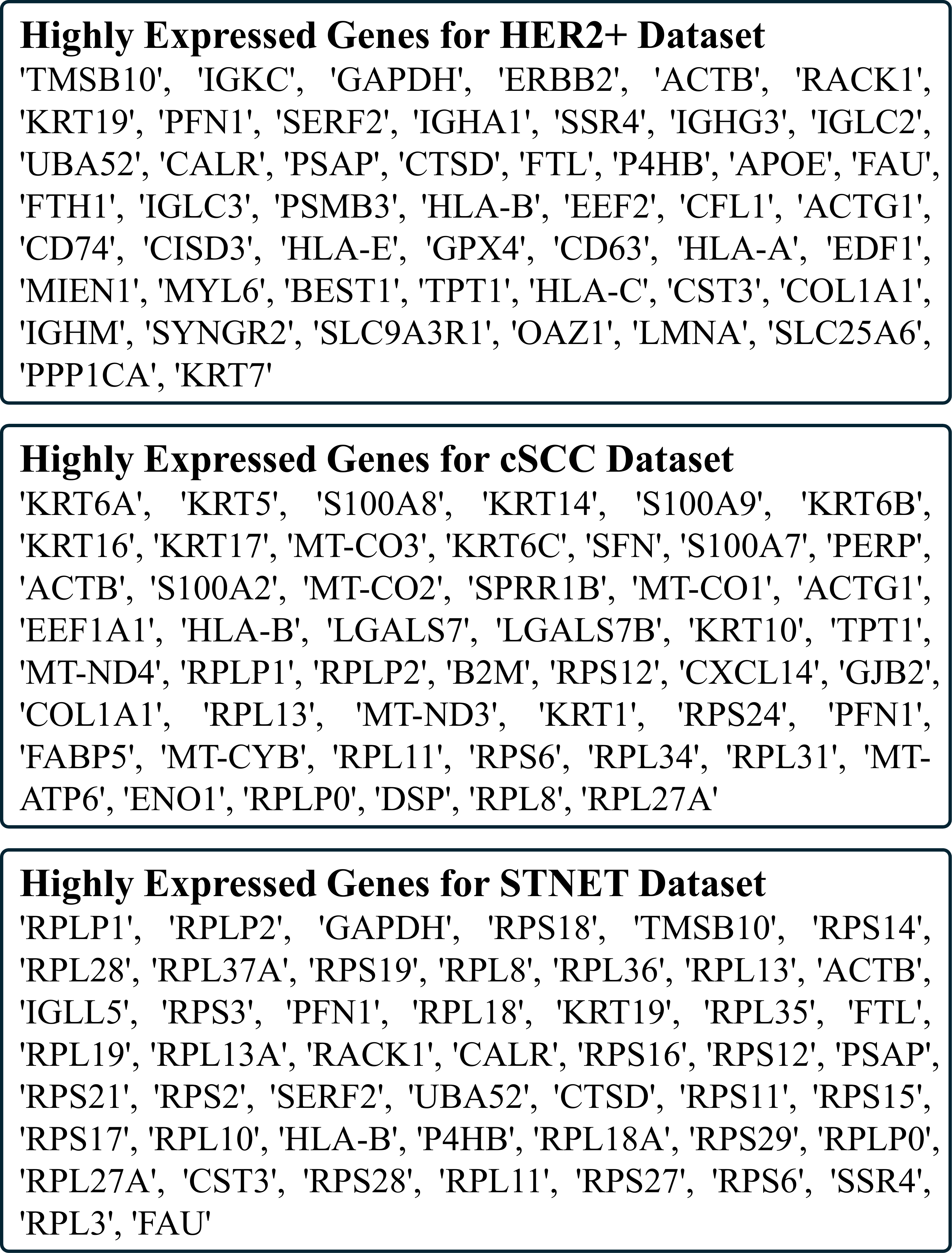}
  \caption{Highly expressed genes for all datasets.}
  \label{fig:heg}
\end{figure}

\begin{figure}[htbp]
  \centering
  \includegraphics[width=0.93\linewidth]{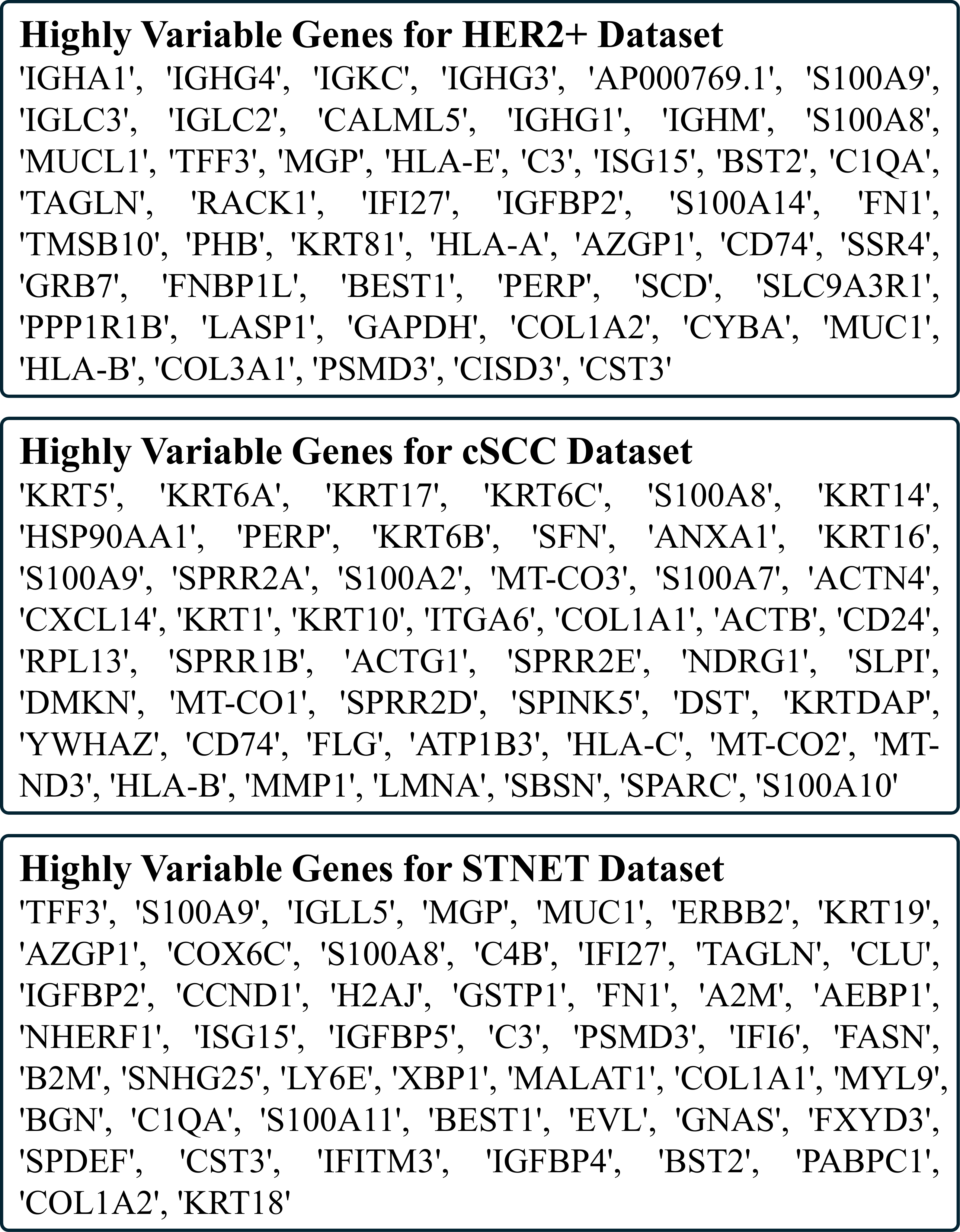}
  \caption{Highly variable genes for all datasets.}
  \label{fig:hvg}
\end{figure}

\section{Preprocessing and Evaluation Protocols}
\label{sec:geneset}

\subsection{Gene Expression Data Preprocessing}
To address the inherent sparsity in spatial transcriptomics data, we first filtered out genes with low variability, following the criteria established in a previous study~\cite{stnet}. Each spot’s gene expression values were then normalized by dividing by the total expression sum, followed by a logarithmic transformation to stabilize variance. To further mitigate experimental noise, we applied a smoothing technique~\cite{stnet}, which averages each spot’s gene expression profile with those of its neighboring spots.

\begin{figure*}[htbp]
  \centering
  \includegraphics[width=\linewidth]{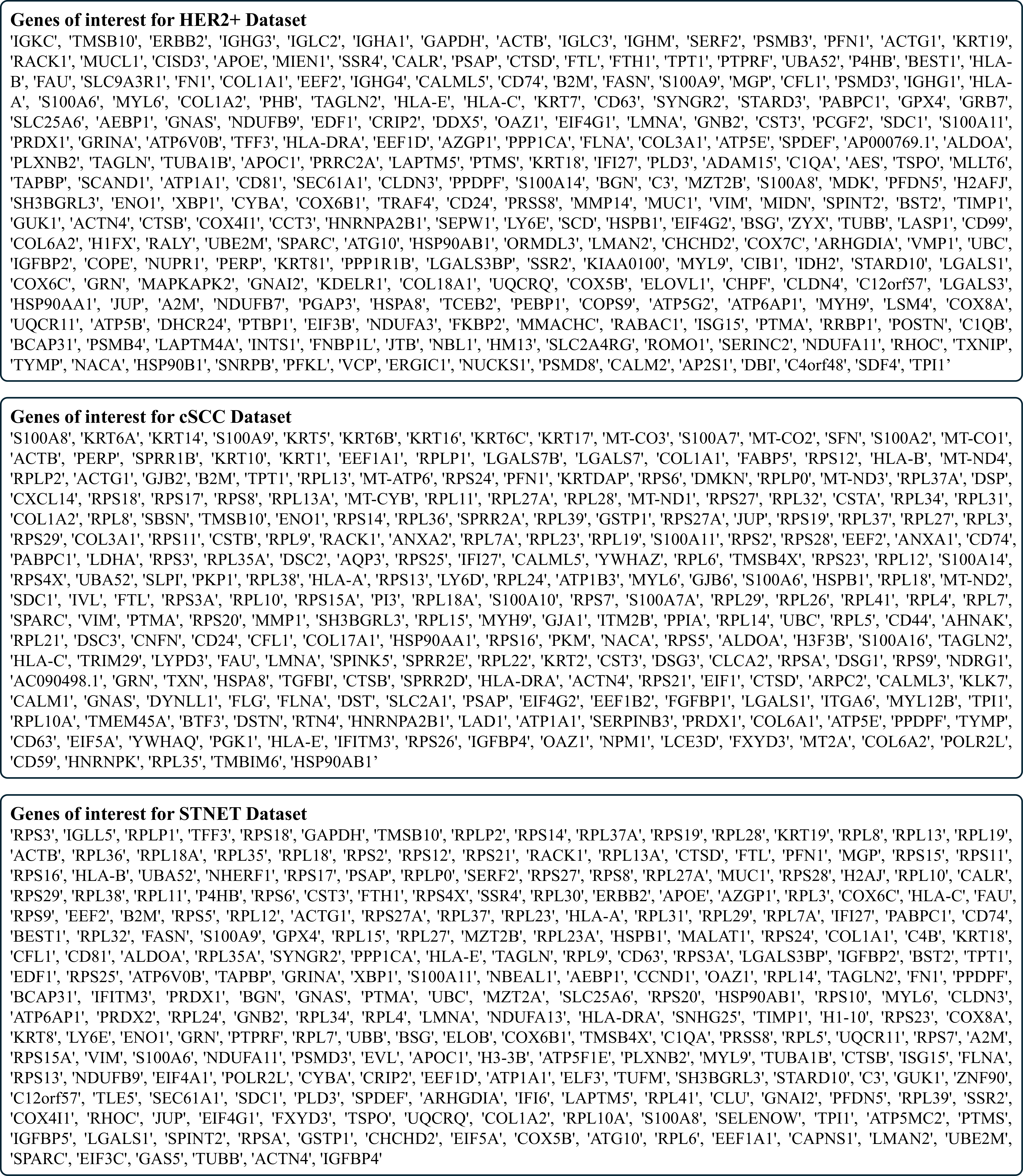}
  \caption{Genes of interest for all datasets.}
  \label{fig:all_gene}
\end{figure*}

\subsection{Evaluation Across Gene Subsets}
To comprehensively assess the effectiveness of our proposed model, we evaluate its performance using the PCC from four complementary perspectives: (1) PCC for Highly Predictive Genes (HPGs) (Figure~\ref{fig:hpg}), (2) PCC for Highly Expressed Genes (HEGs) (Figure~\ref{fig:heg}), (3) PCC for Highly Variable Genes (HVGs) (Figure~\ref{fig:hvg}), and (4) PCC across all genes (Figure~\ref{fig:all_gene}).

While the overall PCC provides a global view of model performance, the three gene-specific subsets offer more targeted insights: (i) HPGs represent genes for which the model achieves the highest predictive accuracy, indicating the model's best-case performance. (ii) HEGs are genes with high average expression, which tend to be more robust to technical noise and biologically significant. (iii) HVGs are genes with high spatial variability, often associated with meaningful biological heterogeneity.

To ensure consistency with prior studies~\cite{bleep,egn,triplex,eggn,mclstexp}, we select the top 50 genes in each category (HPG, HEG, and HVG) based on the corresponding criteria. This stratified evaluation enables a more comprehensive understanding of the model’s performance across genes with distinct biological and statistical properties.

\section*{Acknowledgements}
This work was supported by the Institute of Digital Medicine, City University of Hong Kong, the Hong Kong Innovation and Technology Commission (InnoHK Project CIMDA) and City University of Hong Kong internal grant 7005967.

\clearpage
\bibliography{aaai2026}

\end{document}